\documentclass[12pt,fleqn]{article}
\usepackage[textheight=23.5cm,textwidth=16.5cm]{geometry}
\usepackage{amsmath}
\usepackage{graphicx}
\usepackage[super]{cite}

\usepackage[american]{babel}

\usepackage{dcolumn} %% tables cols aligned at decimal point
\newcolumntype{d}{D{.}{.}{2}}
\newcolumntype{e}{D{.}{.}{3}}
\newcolumntype{f}{D{.}{.}{4}}

\usepackage{threeparttable}

\begin{document}

\begin{center}

{\LARGE\bf
Accurate \emph{ab initio} spin densities
}

{\large 
Katharina Boguslawski$^{\rm a}$,
Konrad H.\ Marti$^{\rm a}$,
\"Ors Legeza$^{\rm b}$,
and\\[0ex] Markus Reiher$^{\rm a,}$\footnote{Author to whom correspondence should be sent; email: markus.reiher@phys.chem.ethz.ch, FAX: +41-44-63-31594, TEL: +41-44-63-34308}
}\\[0ex]

$^{\rm a}$ETH Zurich, Laboratorium f{\"u}r Physikalische Chemie, 
Wolfgang-Pauli-Str.\ 10,\\ [-1ex]
CH-8093 Zurich, Switzerland \\ [0ex]
$^{\rm b}$Wigner Research Centre, P.O Box 49, H-1525 Budapest, Hungary\\[2ex]

\end{center}

\begin{abstract}
We present an approach for the calculation of spin density distributions
for molecules that require very large active spaces for a qualitatively
correct description of their electronic structure. 
Our approach is
based on the density-matrix renormalization group (DMRG) algorithm to calculate the
spin density matrix elements as basic quantity for the spatially resolved spin density distribution.
The spin density matrix elements are directly determined from the second-quantized elementary
operators optimized by the DMRG algorithm.
As an analytic convergence criterion
for the spin density distribution, we employ our recently developed sampling-reconstruction scheme
[{\it J. Chem. Phys.} {\bf 2011}, {\it 134}, 224101]
to build an accurate complete-active-space
configuration-interaction (CASCI) wave function from the optimized
matrix product states.
The spin density matrix elements can then also be determined as an expectation value 
employing the reconstructed wave function expansion.
Furthermore, the explicit reconstruction of a CASCI-type wave function provides insights into
chemically interesting features of the molecule under study such as
the distribution of $\alpha$- and $\beta$-electrons in terms of
Slater determinants, CI coefficients, and natural orbitals.
The methodology is applied to an iron nitrosyl complex which we have identified
as a challenging system for standard approaches
[{\it J. Chem. Theory Comput.} {\bf 2011}, {\it 7}, 2740].
\end{abstract}

\vspace*{-0.5cm}

\begin{tabbing}
Date:   \quad \= April 25, 2012  \\[-1ex]
Status:       \> revised version, published in \textit{J. Chem. Theory Comput.\ {\it 8} {\bf 2012} 1970--1982}
\end{tabbing}

\newpage

%%%%%%%%%%%%%%%%%%%%%%%%%%%%%%%%%%%%%%%%%%%%%%%%%%%%%%%%%%%%%%%%%%%%%%%%%%%%%%%%%%%%%%%%%%%5
\section{Introduction}
%%%%%%%%%%%%%%%%%%%%%%%%%%%%%%%%%%%%%%%%%%%%%%%%%%%%%%%%%%%%%%%%%%%%%%%%%%%%%%%%%%%%%%%%%%%5
In quantum chemistry, the electronic structure of molecules is described by either \emph{ab initio} wave-function
methods or density-functional theory (DFT). For large molecular systems such as transition metal
complexes, however, wave-function based methods are rarely employed due to
the corresponding high computational cost (for counterexamples, see Refs.~\cite{radon,pierloot,roos,Sala2010,Planas2011}). Hence, the
application of DFT became instrumental in theoretical studies of mechanisms in metal-mediated catalysis
\cite{rev_frenking,rev_ziegler,rev_neese,rev_sauer,sauer_jacs,frenking09,frenkingdft,ziegler10,marenadvinorgchem}.
Yet, the treatment of open-shell systems
\cite{markus_fd,markus_chimia_2009,markusbuch} and (near-)degenerate states remains a challenge for DFT \cite{savinbook}.
Failures of approximate exchange--correlation density functionals in predicting properties of
open-shell systems have been traced to the delocalization error and static correlation error
\cite{CohenScience,cohen2012} which are rooted in an inappropriate behavior of the energy with respect to
fractional charges and fractional spins \cite{cohenFractionalSpin}.
In addition to the
difficult prediction of ground states from states of different spin
\cite{tac2001,b3lypstern,markusic2002,topcurchem_spinstates,markuscasida,markusjcc2006,markus_chimia_2009,jchemphys126,
swart_spinstates2008,neese_spinstates}, spin density distributions considerably
depend on the approximate exchange--correlation density functional if transition metal complexes
containing noninnocent ligands are considered \cite{ghosh}. Qualitatively correct spin density distributions
are difficult to obtain within the standard Kohn--Sham formalism that has not been formulated to also produce
accurate spin densities \cite{spindft}.

However, accurate spin densities are desirable for various reasons. (1) In electron paramagnetic resonance spectroscopy
(EPR) \cite{jeschke}, the spin density is the central quantity on which EPR parameters explicitly depend \cite{nmrepr}.
Obviously, reliable spin density distributions are important for an accurate calculation of
EPR properties, but this remains a difficult task to achieve for theoretical chemistry \cite{epr1,epr2,epr3,epr4,epr5}.
(2) The question which approximate exchange--correlation density functional yields sufficiently accurate
spin densities remains inconclusive \cite{ghosh,feno}.
If accurate reference spin density distributions were available, a more detailed analysis of the spin
density distribution in terms of spin density difference plots could be used as a qualitative \emph{and}
quantitative benchmark for the validation of approximate exchange--correlation density functionals. (3) According
to the Hohenberg--Kohn theorem \cite{hohenbergkohn}, the spin density is not needed to calculate the electronic
energy or any other expectation value. However, in open-shell systems it is often introduced as an
additional variable which leads to a spin-DFT formalism \cite{parr} first introduced by von Barth and Hedin
\cite{barth}. In spin-DFT, the spin density becomes a fundamental quantity and reliable
reference spin densities could be used to construct proper approximations to the exact exchange--correlation
density functional.

For accurate spin densities in cases for which a DFT description fails, \emph{ab initio} electron
correlation methods need to be applied. Pierloot \emph{et al.} presented
complete-active-space self-consistent-field (CASSCF) studies for large transition metal complexes
which provided deeper insights into the quality of DFT spin density distributions \cite{pierloot,pierloot3}.
The large molecular size of these systems requires large active orbital spaces, but the standard CASSCF approach
restricts their dimension which represents the most crucial approximation in such calculations \cite{feno}.
It is therefore important to understand whether the spin density is converged with respect to the dimension
of the active orbital spaces used so far. This is a task that is difficult to study within a standard
CAS-type approach.

In general, up to about 18 electrons correlated in 18 spatial orbitals
are computationally feasible for standard CASSCF. These limitations may
restrict the accurate description of electronic structures which could be approved only by enlarging the dimension
of the active orbital  space. Reliable reference spin density distributions for complicated open-shell structures as found,
for instance, in iron complexes with non-innocent ligands require capabilities beyond those of standard correlation methods.

A different approach for the calculation of correlated \emph{ab initio} spin densities for large molecules was recently
presented by Kossmann and Neese \cite{neese_sd} who discussed the performance of orbital-optimized
M$\o$ller--Plesset perturbation theory in calculating hyperfine
coupling constants for atoms and small molecules. In this approach, isotropic hyperfine constants of coupled-cluster singles-doubles
quality could be obtained which could be further improved by applying spin-component scaling.

Here, we pursue a different route for the calculation of \emph{ab initio} spin densities by applying
the density-matrix
renormalization group (DMRG) algorithm. With the DMRG algorithm, introduced by White \cite{white,PhysRevLett.68.3487}
in 1992, much larger active orbital spaces can be considered beyond the limit of, say, 18 electrons correlated in
18 molecular orbitals. It was shown that DMRG is capable of providing accurate
wave functions and energies even for complicated electronic structures (see Refs.~\cite{ors_springer,dmrg_chan,marti2010b,chanreview}
for reviews). Moreover, we first showed that the
DMRG algorithm yields reliable relative electronic energies between different
spin states or isomers of transition metal complexes and clusters for which DMRG was not meant to work and
which are a very challenging task for any other multi-reference quantum chemical
method \cite{marti2008} (see also Ref.~\cite{orbitalordering} for latest results and further references). 
We shall demonstrate in this work that also accurate DMRG spin density distributions
can be determined for very large active orbital spaces.

Recently, we presented a convergence analysis
of the spin density distribution for a small iron nitrosyl model complex [Fe(NO)]$^{2+}$ in a field of point charges,
which demonstrated that
medium-sized active orbital spaces are sufficient for \emph{quantitatively} correct spin densities
\cite{feno}. However, a quantitative analysis that can explore truly large active spaces
is still lacking for this complex, which shall therefore be the target system in this work.
In such cases, DMRG spin densities can be considered as reliable references
which can serve as benchmark results for 
approximate exchange--correlation density functionals.

This work is organized as follows. In section \ref{Sec:theory}, we discuss the spin density
matrix and its spatially resolved counterpart, the spin density distribution,
employing the  formalism of second quantization. Then, we continue with the introduction of DMRG spin densities.
In section \ref{Sec:casci}, we present our approach of approximating the DMRG spin density distribution by
one from a complete-active-space configuration-interaction(CASCI)-type wave function which
allows us to compare DMRG spin densities from calculations with different DMRG parameter sets.
In order to validate our approach, we study the spin densities of a medium-sized
active orbital space in section \ref{Sec:test}. This is then extended by considering up to 29
active orbitals in section \ref{Sec:results}.
Finally, a summary and concluding remarks are given in section \ref{Sec:conclusion}.

%%%%%%%%%%%%%%%%%%%%%%%%%%%%%%%%%%%%%%%%%%%%%%%%%%%%%%%%%%%%%%%%%%%%%%%%%%%%%%%%%%%%%%%%%%%5
\section{Spatially resolved, non-relativistic spin densities \label{Sec:theory}}
%%%%%%%%%%%%%%%%%%%%%%%%%%%%%%%%%%%%%%%%%%%%%%%%%%%%%%%%%%%%%%%%%%%%%%%%%%%%%%%%%%%%%%%%%%%5
Since DMRG is based on the second quantized formalism, we briefly discuss
how the spin density in spatial coordinates can be written in second quantization.
In first quantization, the operator for the spin density reads
\begin{equation}\label{Eq:sd_firstquant}
\hat{\delta}^{\rm spin}({\bf r}) = \sum_{i=1}^N \delta({\bf r - r}_i)\hat{s}_{z,i},
\end{equation}
where $\hat{s}_{z,i}$ is the $z$ component of the one-electron spin operator,
${\bf r}_i$ is the spatial coordinate of electron $i$, and $N$ is the total number
of electrons in the system. Applying an orbital basis, the corresponding operator expression in second
quantization is given by
\begin{align}\label{Eq_sd_operator}
 \hat{\delta}^{\rm spin}({\bf r}) &= \frac{1}{2}\sum_{p,q}\phi_p^\ast({\bf r})
\phi_q({\bf r}) \left( a^\dagger_{p\alpha}a_{q\alpha}-a^\dagger_{p\beta}a_{q\beta}\right) \nonumber\\
&= \sum_{p,q}\phi_p^\ast({\bf r})\phi_q({\bf r}) \hat{ T}_{pq},
\end{align}
where $p,q$ run over the total orbital basis \{$\phi_i$\} with $\phi_i({\bf r})$ representing the spatial
part of a spin orbital. The operators $a^\dagger_{i\sigma}$ and $a_{i\sigma}$ are the creation and annihilation
operators, respectively, for an electron of spin $\sigma$ in orbital $i$. In Eq.\,(\ref{Eq_sd_operator}), the spin density operator
$ \hat{\delta}^{\rm spin}({\bf r}) $ is defined in terms of the spin tensor excitation operators
\begin{equation}\label{Eq:tpq}
\hat{ T}_{pq}=\frac{1}{2} (a^\dagger_{p\alpha}a_{q\alpha}-a^\dagger_{p\beta}a_{q\beta})
\end{equation}
in the orbital basis (see Ref.~\cite{jorgensen} for details). The spatially resolved spin density
$\rho^{\rm spin}({\bf r})$ is calculated as the expectation value of $\hat{\delta}^{\rm spin}({\bf r})$,
\begin{align}\label{Eq:sd_def}
\rho^{\rm spin}({\bf r}) &= \langle \Psi_M \vert \hat{\delta}^{\rm spin}({\bf r}) \vert \Psi_M \rangle \nonumber \\
&=  \sum_{pq}\phi_p^\ast({\bf r})\phi_q({\bf r}) \langle \Psi_M \vert  \hat{ T}_{pq} \vert \Psi_M \rangle,
\end{align}
where $\vert \Psi_M\rangle$ represents some normalized reference state
\begin{equation}
\vert \Psi_M \rangle = \sum_{\bf \{n\}} C^{(M)}_{\bf \{n\}}\vert {\bf n} \rangle.
\end{equation}
$\vert {\bf n}  \rangle = \vert n_1n_2\ldots n_k\rangle$ is an occupation number vector with elements $n_p
\in \{0,1\}$. ${\bf \{n\}}$ represents the set of all occupation number vectors constructed from $k$
one-particle states. The expectation value
on the right hand side of Eq.\,(\ref{Eq:sd_def}) is a spin density matrix element ${ T}^{\rm (M)}_{pq}$,
\begin{align}\label{Eq:sd_matrix_def}
{ T}^{\rm (M)}_{pq} =  \langle \Psi_M \vert \hat{ T}_{pq} \vert \Psi_M \rangle = \frac{1}{2}
\langle \Psi_M \vert a^\dagger_{p\alpha}a_{q\alpha} - a^\dagger_{p\beta}a_{q\beta} \vert \Psi_M \rangle.
\end{align}

%%%%%%%%%%%%%%%%%%%%%%%%%%%%%%%%%%%%%%%%%%%%%%%%%%%%%%%%%%%%%%%%%%%%%%%%%%%%%%%%%%%%%%%%%%%5
\subsection{DMRG spin densities from second-quantized elementary operators\label{Sec:dmrg_sd}}
%%%%%%%%%%%%%%%%%%%%%%%%%%%%%%%%%%%%%%%%%%%%%%%%%%%%%%%%%%%%%%%%%%%%%%%%%%%%%%%%%%%%%%%%%%%5
If the reference state $|\Psi_M\rangle$ is a DMRG wave function in Eq.\,(\ref{Eq:sd_matrix_def}),
the corresponding DMRG spin density matrix elements ${T}^{\rm (M[DMRG])}_{pq}$ are obtained.
The matrix representations of the creation and annihilation operators are available in every step of the
DMRG algorithm and each spin density matrix element can thus be easily determined.

The operator $a^\dagger_{p\sigma}a_{q\sigma}$ in its matrix representation is calculated as a tensor product
for which we have to distinguish two different cases. The molecular orbitals $p$ and
$q$ are defined either (i) on the same or (ii) on different subsystems of the DMRG partitioning of the active
orbital space into the active (sub)system, its environment (the complementary subsystem), and one or
two explicitly treated orbitals in between.
While the former case is straightforward to handle, for an operator expression in
the latter case, however, we need to build operators for the superblock where all three subsystems,
i.e., the active subsystem, the exactly represented sites and the environment, are combined as tensor products.

To illustrate this concept, let us consider two operators $a_1$ and $a_2$ defined on three different subspaces
$\mathcal {\tilde{F}}_1$, $\mathcal {\tilde{F}}_2$, and $\mathcal {\tilde{F}}_3$.
Then, the combined subspace $\mathcal {\tilde F}$ is defined as $ \mathcal {\tilde F}=
\mathcal {\tilde{F}}_1\otimes \mathcal {\tilde{F}}_2 \otimes  \mathcal {\tilde{F}}_3$, where 
$\mathcal {\tilde F}$ as well as $\mathcal {\tilde{F}}_1$, $\mathcal {\tilde F}_2$ and
$\mathcal {\tilde{F}}_3$ are all subspaces of the $N$-particle Fock space
$\mathcal {\tilde{F}}_N$. For instance, the operator expressions for the combined subspace are given by
\begin{align}
a_1^{\mathcal {\tilde{F}}}&: \quad a_1 \otimes {\bf 1}_{\mathcal {\tilde{F}}_2}\otimes{\bf 1}_{\mathcal {\tilde{F}}_3}\\
a_2^{\mathcal {\tilde{F}}}&: \quad {\bf A}_{\mathcal {\tilde{F}}_1}\otimes a_2 \otimes {\bf 1}_{\mathcal {\tilde{F}}_3},
\end{align}
where ${\bf A}_{\mathcal {\tilde{F}}_i}$ is the anticommutation matrix of the corresponding subspace ${\mathcal {\tilde{F}}_i}$.
For the product of two operators we obtain
\begin{equation}
a_1^{\mathcal {\tilde{F}}}\cdot a_2^{\mathcal {\tilde{F}}} = \left(a_1 \cdot {\bf A}_{\mathcal {\tilde{F}}_1} \right)\otimes 
\left( a_2\right)\otimes \left( {\bf 1}_{\mathcal {\tilde{F}}_3}\right),
\end{equation}
where we have used the mixed-product property for the right-hand side of the above equation which mixes the ordinary matrix
product with the tensor product. All remaining operator products
can be derived in a similar way. After the spin density matrix is determined, the spatially resolved
spin density distribution can be calculated from Eq.\,(\ref{Eq:sd_def}). If the wave
function is real, the spin density matrix is symmetric and the calculation can
be speed up by calculating the upper triangular part of the spin density matrix
only.

%%%%%%%%%%%%%%%%%%%%%%%%%%%%%%%%%%%%%%%%%%%%%%%%%%%%%%%%%%%%%%%%%%%%%%%%%%3yy%%%%%%%%%%%%%%%%%5
\subsection{Spin density from a reduced dimensional CASCI-type wave function\label{Sec:casci}}
%%%%%%%%%%%%%%%%%%%%%%%%%%%%%%%%%%%%%%%%%%%%%%%%%%%%%%%%%%%%%%%%%%%%%%%%%%%%%%%%%%%%%%%%%%%5
Since CI vectors are in general
sparse \cite{greer,Mitrushenkov,srcas_jcp}---if contributions below a predefined
threshold are neglected---CASCI-type wave functions can be efficiently and accurately projected onto a smaller set of Slater
determinants, which only represent the most important contributions to the wave function expansion.
We recently reported the sampling-reconstruction algorithm for CASCI-type wave functions defined in a complete
active orbital space from a previously optimized DMRG wave function (SRCAS algorithm) \cite{srcas_jcp}. 
An approximate CASCI-type expansion $|\tilde{\Psi}_M\rangle$ for any wave function
$|\Psi_M\rangle$ consisting of $k$ one-particle states can thus be written as
\begin{equation}\label{Eq:wfct_expansion}
|\tilde{\Psi}_M \rangle=\sum_{\{\tilde{\bf n}\}} C^{(M)}_{\{\tilde{\bf n}\}} \vert \tilde{\bf n} \rangle,
\end{equation} 
where the sum runs over all occupation number vectors $\tilde{\bf n}$ living in
the sampled subspace of the total many-particle Hilbert space. Using Eq.\,(\ref{Eq:sd_matrix_def}), we can calculate 
the spin density matrix by substituting the reference state $|\Psi_M\rangle$ with
the approximate state $|\tilde{\Psi}_M\rangle$,
\begin{align}\label{Eq:sd_matrix_approx}
{ T}^{\rm (M[SRCAS])}_{pq} &= \langle \tilde{\Psi}_M \vert \hat{T}_{pq} \vert \tilde{\Psi}_M \rangle = \frac{1}{2}
\langle \tilde{\Psi}_M \vert a^\dagger_{p\alpha}a_{q\alpha} - a^\dagger_{p\beta}a_{q\beta} \vert \tilde{\Psi}_M\rangle \nonumber\\
&=\frac{1}{2}\sum_{\{\tilde{\bf n}\}, \{\widetilde{\bf m}\}}  C_{\{\tilde{\bf n}\}}^{(M)*}  C^{(M)}_{\{\tilde{\bf m}\}}
  \langle \tilde{\bf n} \vert a^\dagger_{p\alpha}a_{q\alpha} - a^\dagger_{p\beta}a_{q\beta} \vert \widetilde{\bf m}
\rangle.
\end{align}
Since the occupation number vectors are orthonormal to each other, the expectation value on the right hand
side of Eq.\,(\ref{Eq:sd_matrix_approx}) can be easily evaluated and we obtain
\begin{align}
{ T}^{\rm (M[SRCAS])}_{pq} &= \frac{1}{2}\sum_{\{\tilde{\bf n}\}, \{\widetilde{\bf m}\}}  C_{\{\tilde{\bf n}\}}^{(M)*}  C^{(M)}_{\{\widetilde{\bf m}\}}
  \left(\epsilon_{pq\alpha} \delta_{\tilde{\bf n}_{\widehat{p\alpha}} \widetilde{\bf m}_{\widehat{q\alpha}}} -
 \epsilon_{pq\beta} \delta_{\tilde{\bf n}_{\widehat{p\beta}} \widetilde{\bf m}_{\widehat{q\beta}}} \right),
\end{align}
where $\tilde{\bf n}_{\widehat{p\sigma}}$ represents the occupation number
vector where orbital $p$ lacks one electron with
$\sigma$ spin. Furthermore, we introduced a phase factor $\epsilon_\sigma$
to account for the annihilation operations of $a^\dagger_{p\sigma}$ acting on
the bra-state and $a_{q\sigma}$ acting on the ket-state.

Based on this approximate expression for the spin density matrix
we can determine spin density distributions for subspaces of the many-particle Hilbert space of different
dimensions and study the sensitivity of the spin density distribution to the number of active-system states 
in DMRG calculations.  

%%%%%%%%%%%%%%%%%%%%%%%%%%%%%%%%%%%%%%%%%%%%%%%%%%%%%%%%%%%%%%%%%%%%%%%%%%3yy%%%%%%%%%%%%%%%%%5
\subsection{Measures for spin density comparisons\label{Sec:errors}}
%%%%%%%%%%%%%%%%%%%%%%%%%%%%%%%%%%%%%%%%%%%%%%%%%%%%%%%%%%%%%%%%%%%%%%%%%%%%%%%%%%%%%%%%%%%5
For various reasons, we need suitable measures to assess the similarity of different spin densities.
For instance, such a measure would be required to assess the accuracy of a given spin density compared to a reference
spin density. 

Monitoring the evolution of the spin density for an increasing number of active-system states \cite{gerrit_sd}
can illustrate the convergence behavior of the spin density distribution with respect to the number
of active-system states $m$. Isosurface plots of the
difference in spin density distributions for calculations with different $m$-values can only serve as a qualitative
convergence measure. As quantitative measures, however, we introduce two distances which quantify how far
two spin densities are apart from each other. Both distance measures are defined with
the absolute error in the spin density difference distribution. The accumulated absolute error $\Delta_{\rm abs}$
is given by
\begin{equation}\label{deltaabs}
 \Delta_{\rm abs} = \int |\rho^{\rm spin}_{1}({\bf r}) - \rho^{\rm spin}_{2}({\bf r}) | {\rm d}{\bf r},
\end{equation}
and the root-square error $\Delta_{\rm rs}$ reads
\begin{equation}\label{deltasq}
 \Delta_{\rm rs} = \sqrt{\int |\rho^{\rm spin}_{1}({\bf r}) - \rho^{\rm spin}_{2}({\bf r}) |^2 {\rm d}{\bf r}},
\end{equation}
where $\rho^{\rm spin}_{i}({\bf r})$ refers to the spin density distribution corresponding to some calculation $i$,
e.g., to a CASSCF or DMRG spin density when different chemical methods are compared, or to some parameter sets if
different spin densities are determined with the same method.
If two spin densities $\rho^{\rm spin}_{i}({\bf r})$ and $\rho^{\rm spin}_{j}({\bf r})$ are similar, both $\Delta_{\rm abs}$ and
$\Delta_{\rm rs}$ approach zero.
For accurate \emph{ab initio} spin densities, we shall require both error measures to
be smaller than 0.005 ($\Delta_{\rm abs}$) or 0.001 ($\Delta_{\rm rs}$), respectively (in view of the results
discussed in section \ref{Sec:single_ref}).

A different similarity measure can be applied by employing directly the knowledge of the reconstructed
CASCI-type wave function expansion. This procedure
relies on the closeness measure of two quantum states, namely the quantum fidelity \cite{fidelity1,fidelity2}.
The importance and potential application of the quantum fidelity within the DMRG framework was first
discussed by some of us \cite{legeza_dbss} in the context of quantum error correction and was also utilized
in our SRCAS approach \cite{srcas_jcp}.
Two CASCI-type wave function expansions reconstructed for different numbers of DMRG
active-system states, $m_1$ and $m_2$, can be explicitly compared by calculating their quantum fidelity
\begin{equation}
F_{m_1,m_2}=|\langle \tilde{\Psi}^{(m_1)}_M | \tilde{\Psi}^{(m_2)}_M \rangle|^2
\end{equation}
as an overlap measure.

%%%%%%%%%%%%%%%%%%%%%%%%%%%%%%%%%%%%%%%%%%%%%%%%%%%%%%%%%%%%%%%%%%%%%%%%%%%%%%%%%%%%%%%%%%%5
\section{A noninnocent model system \label{Sec:test}}
%%%%%%%%%%%%%%%%%%%%%%%%%%%%%%%%%%%%%%%%%%%%%%%%%%%%%%%%%%%%%%%%%%%%%%%%%%%%%%%%%%%%%%%%%%%5
In a previous study, we reported DFT and CASSCF spin density distributions
in iron nitrosyl complexes as well as for the [Fe(NO)]$^{2+}$ molecule embedded in a square-planar
field of point-charges to emulate the one-electron states of the full complexes \cite{feno}.
Since DFT spin densities of iron nitrosyl complexes remain ambiguous, we choose the small
[Fe(NO)]$^{2+}$ molecule in its doublet state for our analysis here. The point charges
facilitate a dynamic change of the character of the electronic wave function by shortening
the distances $d_{\rm pc}$ of the point charges to
the metal center. Depending on this distance $d_{\rm pc}$, both single-reference and multi-reference
situations can be created for [Fe(NO)]$^{2+}$.
When the four point charges are located at a distance of 1.131 \AA{} from the iron atom, the electronic
structure of the [Fe(NO)]$^{2+}$ molecule represents a single-reference problem,
while for $d_{\rm pc}=0.598$ \AA{} a multi-reference case is generated.

The [Fe(NO)]$^{2+}$ structure features a Fe--N bond length of 1.707 \AA{}
and a N--O bond distance of 1.177 \AA{} with a Fe--N--O angle of 146$^\circ$. The four negative point charges
of $-$0.5 $e$ each are located as depicted in Figure~\ref{fig:feno_dens}(e).
Due to the small size of the [Fe(NO)]$^{2+}$ molecule, we can efficiently study the dependence of the
spin density distribution on different DMRG parameter sets such as the number of DMRG active-system states $m$.
Thereby, we are able to define appropriate convergence measures for the spin density in order to reach
a predefined accuracy.

%%%%%%%%%%%%%%%%%%%%%%%%%%%%%%%%%%%%%%%%%%%%%%%%%%%%%%%%%%%%%%%%%%%%%%%%%%%%%%%%%%%%%%%%%%%5
\subsection{The single-reference case \label{Sec:single_ref}}
%%%%%%%%%%%%%%%%%%%%%%%%%%%%%%%%%%%%%%%%%%%%%%%%%%%%%%%%%%%%%%%%%%%%%%%%%%%%%%%%%%%%%%%%%%%5
As already discussed in great detail in Ref.~\cite{feno}, the minimal active orbital space for [Fe(NO)]$^{2+}$ with
$d_{\rm pc}=1.131$ \AA{} comprises seven electrons correlated in seven orbitals
for qualitatively reliable spin density distributions. It consists of Fe $3d$-
($d_{xy}$, $d_{yz}$, $d_{xz}$, $d_{x^2-y^2}$ and $d_{z^2}$) and both NO $\pi^*$-orbitals.
As orbital basis in our DMRG calculations, the natural orbitals from a CAS(7,7)SCF calculation
performed with the \textsc{Molpro} program package \cite{molpro} using Dunning's cc-pVTZ basis set
for all atoms \cite{dunning,dunning2} were taken.
The one-electron and two-electron integrals in the natural orbital basis
were also calculated with the \textsc{Molpro} program package \cite{molpro}. All DMRG calculations reported in this section \ref{Sec:test}
were carried out with the Zurich DMRG program \cite{dmrg_new}. Random noise was added to the density matrix
in order to force the mixing of configurations that would have not been captured otherwise if the number
of active-system states $m$ is too small \cite{channoise,dorandonoise}.

\begin{figure}[h]
\centering
\includegraphics[width=0.8\linewidth]{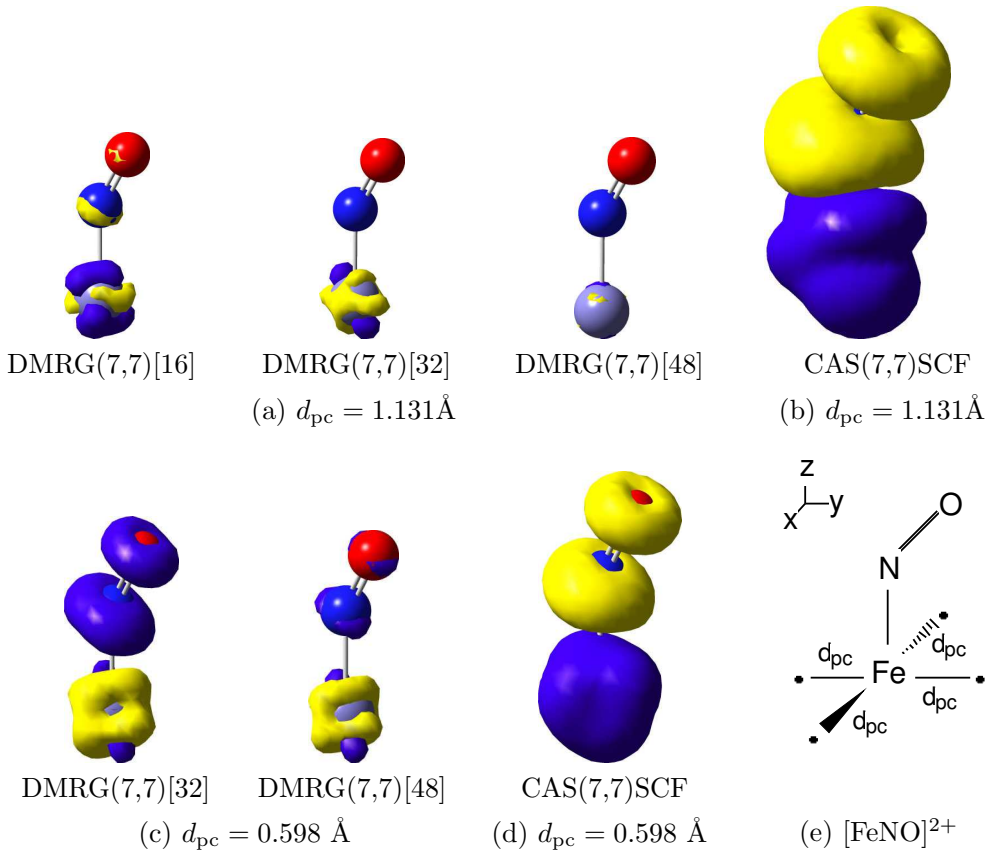}
\caption{(a) and (c): Spin density difference plots for DMRG(7,7)[$m$] spin densities calculated for a different number of
DMRG active-system states $m$ with respect to the CAS(7,7)SCF reference spin densities shown in (b) and (d), respectively, for
[Fe(NO)]$^{2+}$ in a quadratic-planar point-charge field. Two different distances are considered,
namely $d_{\rm pc} = 1.131$ \AA{}, (a) and (b), and $d_{\rm pc}=0.598$ \AA{}, (c) and (d).
For both distances, the spin density could be perfectly reproduced in a DMRG(7,7)[$64$] calculation and is
therefore not depicted here. An isosurface value of 0.0003 in (a), (b) and 0.003 in (c), (d), respectively, is chosen.
(e): Structure of [Fe(NO)]$^{2+}$ with the four point charges of $-0.5$ e.
}\label{fig:feno_dens}
\end{figure}

We performed DMRG calculations for different numbers of DMRG active-system states $m$ abbreviated
by DMRG($x$,$y$)[$m$] where $x$ corresponds to the number of active electrons and $y$ is the number of active orbitals
for $m$ renormalized active-system states.
Starting with $m=16$, $m$ is further increased to 32 and 48 until the CAS(7,7)SCF reference energy is reproduced
for $m=64$ active-system states (see Table \ref{tab:energiesfeno}).
Note that the number of active-system states needed to reproduce the CASSCF result is very small in this case. This can be
explained employing concepts of quantum information theory in section~\ref{Sec:results}. The DMRG calculations reported in this
section do not employ these concepts to enforce better convergence. This decision is deliberately taken in order to produce
nonconverged low-$m$ results to compare with the $m=64$ calculation.
We should note that this artifact could be cured by the dynamical block state selection (DBSS) procedure \cite{legeza_dbss2,legeza_dbss3},
while the strong dependence on small $m$-values and the convergence to local minima can be overcome
by applying the configuration interaction-based dynamically extended active space (CI-DEAS) procedure \cite{legeza_dbss}.
\begin{table}[h]
\caption{Ground state energy for [FeNO]$^{2+}$ surrounded by four point charges at two different
distance sets $d_{\rm pc}$ in Hartree atomic units
for CAS(7,7)SCF and DMRG(7,7)[$m$] calculations for different numbers of DMRG active-system states $m$.}\label{tab:energiesfeno}
{
\begin{center}
\begin{tabular}{lc|lc}\hline \hline
\multicolumn{2}{c|}{$d_{\rm pc}=1.131$ \AA{}} & \multicolumn{2}{c}{$d_{\rm pc}=0.598$ \AA{}}\\ \hline
Method & $E$/Hartree & Method & $E$/Hartree \\\hline
 HF		  &  $-$1392.844 043 & HF	     &  $-$1396.821 220 \\
 CAS(7,7)SCF	  &  $-$1392.887 247 & CAS(7,7)SCF   &  $-$1396.858 313 \\ \hline
 DMRG(7,7)[$16$]  &  $-$1392.881 067 & DMRG(7,7)[$16$]    &  $-$1396.762 709 \\
 DMRG(7,7)[$32$]  &  $-$1392.885 462 & DMRG(7,7)[$32$]    &  $-$1396.818 651 \\
 DMRG(7,7)[$48$]  &  $-$1392.886 893 & DMRG(7,7)[$48$]    &  $-$1396.840 018 \\
 DMRG(7,7)[$64$]  &  $-$1392.887 247 & DMRG(7,7)[$64$]    &  $-$1396.858 313    
\\ \hline
\hline
\end{tabular}
\end{center}
}
\end{table}
The spin density distributions for our four DMRG calculations ($m=16,32,48,64$)
are shown in Figure \ref{fig:feno_dens}(a) and were determined as discussed in section
\ref{Sec:dmrg_sd}. To emphasize the dependence on $m$,
the corresponding spin density difference plots with respect to the CAS(7,7)SCF reference spin density
distribution are displayed. Note that all isosurface plots are shown for the same isosurface value of
0.0003. All DMRG calculation yield qualitatively similar spin density distribution, only minor quantitative
differences can be observed.
The CAS(7,7)SCF reference spin density can be perfectly well reproduced for $m=64$ DMRG active-system states
and is, hence, not shown in Figure \ref{fig:feno_dens}(a).
To calculate approximate spin density distributions from reconstructed CASCI-type wave functions, we first have to sample
the most important configurations of the $N$-particle Hilbert space. For this purpose, we applied our
SRCAS method \cite{srcas_jcp}. Due to the small size of the active
space, the $N$-particle Hilbert space is spanned by only 1225 Slater determinants and all corresponding CI
coefficients can be determined directly from the CASSCF reference calculation. 
In general, similar CI coefficients are obtained for all DMRG calculations and the CASSCF reference, i.e.,
similar wave functions are converged resulting in small differences in the spin density distributions.
The distribution of the CI coefficients is depicted in Figure~1 of the Supporting Information.

Spin density distributions determined for different sampled subspaces of the $N$-particle Hilbert
are in good agreement with the corresponding DMRG spin density. Note that the sampled subspaces are
defined by the threshold value of the completeness measure (COM) introduced
in Ref.~\cite{srcas_jcp} with COM $= (1 - \sum_I C_{I}^2)$, where $I$ runs over all sampled configurations
with CI coefficients ${C}_{I}$.
In general, threshold values of 0.01 to 0.001 turned out to be sufficient for obtaining
quantitatively reliable spin densities in this single-reference case. The corresponding isosurface plots and excitation
histograms with respect to the COM are summarized in the Supporting Information.

The spin density difference plots in Figure \ref{fig:feno_dens}(a) illustrate the convergence of the
spin density distribution with respect to the number of DMRG active-system states $m$. The absolute error $\Delta_{\rm abs}$
and the root-square error $\Delta_{\rm rs}$ of the spin density difference distributions provide a quantitive
measure for the accuracy (see Table \ref{tab:error_dens_casscf_small}).
The differences in the spin densities calculated for $m=48$ DMRG active-system states is small compared to the CAS(7,7)SCF
reference.
For 48 active-system states upwards, both $\Delta_{\rm abs}$ and $\Delta_{\rm rs}$ are below their threshold values
given in section \ref{Sec:errors}.
The set of quantum fidelity measures $F_{m_i,m_{i+1}}$ for our four DMRG
calculations with $m_i \in \{16,32,48,64\}$ is \{0.980000,0.994395,0.999012\}. Increasing $m$ from 48 to 64 DMRG active-system states
corresponds to $F_{48,64}=0.999012$ which illustrates the similarity of both DMRG wave functions and results in
reliable spin density distributions for $m \geq 48$.

\begin{table}[h]
\caption{The absolute error $\Delta_{\rm abs}$ and the root-square error $\Delta_{\rm rs}$ of the DMRG(7,7)[$m$]
spin densities with respect to the CAS(7,7)SCF reference for [FeNO]$^{2+}$ surrounded by four point charges at two different
distance sets $d_{\rm pc}$ employing different numbers of DMRG active-system states $m$.
}\label{tab:error_dens_casscf_small}
{
\begin{center}
\begin{tabular}{l|cc|cc}\hline \hline
 Method & \multicolumn{2}{c|}{$d_{\rm pc} = 1.131$ \AA}   & \multicolumn{2}{c}{$d_{\rm pc} = 0.598$ \AA} \\\cline{2-5}     
 & $\Delta_{\rm abs}$ & $\Delta_{\rm rs}$ & $\Delta_{\rm abs}$ & $\Delta_{\rm rs}$ \\ \hline
DMRG(7,7)[16] & 0.007678 & 0.002147 & 0.213543  & 0.052168 \\
DMRG(7,7)[32] & 0.004392 & 0.001285 & 0.221198  & 0.052144 \\
DMRG(7,7)[48] & 0.001397 & 0.000412 & 0.081631  & 0.020418 \\
DMRG(7,7)[64] & 1.34$\cdot10^{-5}$  & 5.40$\cdot10^{-6}$ & 9.69$\cdot10^{-6}$ & 3.66$\cdot10^{-6}$ 
\\ \hline
\hline
\end{tabular}
\end{center}
}
\end{table}

%%%%%%%%%%%%%%%%%%%%%%%%%%%%%%%%%%%%%%%%%%%%%%%%%%%%%%%%%%%%%%%%%%%%%%%%%%%%%%%%%%%%%%%%%%%5
\subsection{The multi-reference case \label{Sec:test_multi}}
%%%%%%%%%%%%%%%%%%%%%%%%%%%%%%%%%%%%%%%%%%%%%%%%%%%%%%%%%%%%%%%%%%%%%%%%%%%%%%%%%%%%%%%%%%%5
A multi-reference character of the
[Fe(NO)]$^{2+}$ molecule can be induced by decreasing the distances of the point charges to the iron atom.
In the squeezed model complex, the point charges are placed at a distance of $d_{\rm pc}= 0.598$ \AA{} from the iron center
in the same configuration as before.
Similar to the single-reference problem, the minimum active orbital space considered here comprises seven electrons
correlated in seven orbitals. Yet, it consists of four Fe $d_{xy}$, $d_{yz}$, $d_{xz}$ and $d_{z^2}$ (the $d_{x^2-y^2}$
is excluded due to the compressed point charge environment), two
NO $\pi^*$ and one NO $\sigma$ orbital which interacts with the Fe $d_{z^2}$ orbital.
Again, the natural orbitals from a CAS(7,7)SCF calculation were taken as orbital basis in our DMRG calculations
and determined with the \textsc{Molpro} program package \cite{molpro} using Dunning's cc-pVTZ basis set
for all atoms \cite{dunning,dunning2}. The calculation of the one-electron and two-electron integrals in this natural
orbital basis was also performed with the \textsc{Molpro} program package \cite{molpro}.
All DMRG calculations were carried out with the Zurich DMRG
program \cite{dmrg_new}.
As before, we performed DMRG calculations for four different numbers of DMRG active-system states $m$.
Starting with $m=16$, $m$ is further increased to 32 and 48 until the CAS(7,7)SCF reference energy is obtained
for $m=64$ active-system states (see Table \ref{tab:energiesfeno}).
The small-$m$ calculations are designed not to reproduce the CAS(7,7)SCF reference for this analysis.
Note, however, that a small number of
active-system states was needed to reproduce the CASSCF result as observed in the single-reference problem.

In Figure \ref{fig:feno_dens}(d), the CAS(7,7)SCF spin density distribution is shown which is
taken as the reference distribution, while Figure~\ref{fig:feno_dens}(c) illustrates the spatially resolved
differences in the DMRG(7,7)[$m$] and CAS(7,7)SCF spin density distributions. Note that the same
isosurface value of 0.003 was chosen for all spin densities shown.
For small $m$-values, qualitatively different spin density distributions are obtained. 
The $\beta$-electron density around the nitrosyl ligand is underestimated and a dumbbell-shaped
$\beta$-electron density is obtained in contrast to the cylindric shape of the reference $\beta$-electron density.
The $\alpha$-electron density around the Fe atom is underestimated. Increasing $m$ to 48 results in a cylindric
$\beta$-electron density around the NO-ligand which differs only little from the reference spin density.
The spin density can be exactly reproduced for $m=64$ active-system state for which also the CAS(7,7)SCF reference energy is obtained.
The convergence properties of the DMRG(7,7)[$m$] spin density with respect to $m$ can be quantified by the
$\Delta_{\rm abs}$- and $\Delta_{\rm rs}$-values where significantly large values ($>0.005$ and $>0.001$, respectively)
are obtained for spin density distributions determined in small-$m$ calculations (see Table \ref{tab:error_dens_casscf_small}).

\begin{figure}[h]
\centering
\includegraphics[width=0.8\linewidth]{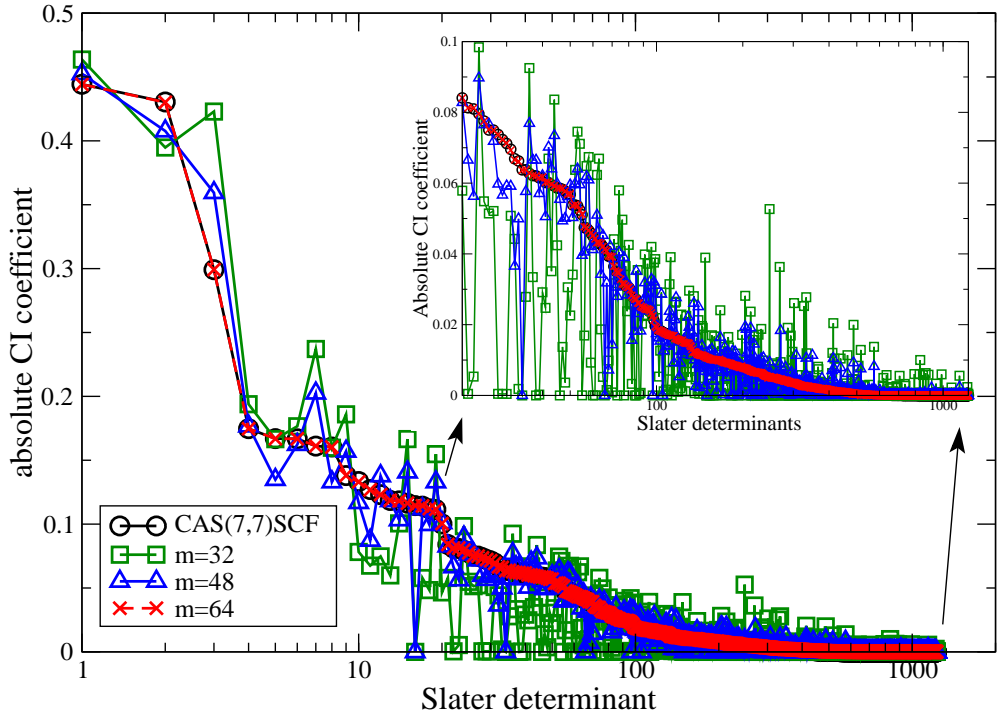}
\caption{Distribution of the absolute value of the CI coefficients corresponding to the Slater determinants
in the DMRG(7,7)[$m$] calculations with different renormalized active-system states $m$ and in the CAS(7,7)SCF reference calculation
for [Fe(NO)]$^{2+}$ surrounded by four point charges at a distance of $d_{\rm pc} = 0.598$ \AA{} from the iron atom.
All Slater determinants are ordered according to the CI weights of the CAS(7,7)SCF calculation.
}\label{fig:feno_cigraph}
\end{figure}

In Figure \ref{fig:feno_cigraph}, the distribution of CI coefficients for the DMRG and CASSCF wave functions is shown. 
Since only the position of the point charges has been modified,
the $N$-particle Hilbert space remains spanned by 1225 Slater determinants and all corresponding CI
coefficients can be determined directly from the CASSCF reference calculation as in the single-reference case.
Similar CI coefficients are obtained for the
DMRG(7,7)[64] calculation and the CAS(7,7)SCF reference, i.e., similar wave functions are converged.
However, significantly different CI coefficients are obtained---as expected---for smaller $m$-values.
In particular, the deviations are most significant for configurations corresponding
to the largest CI weights. Additional information
on the distribution of CI coefficients can be found in the Supporting information.

\begin{figure}[h]
\centering
\includegraphics[width=1.0\linewidth]{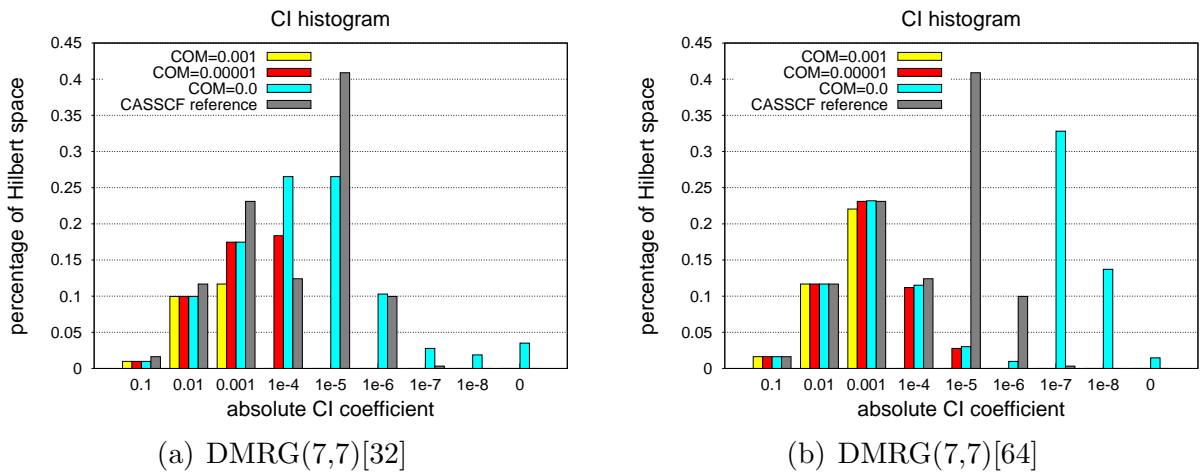}
\caption{CI histogram of the absolute values of the CI coefficients for the Slater determinants
for reconstructed CASCI-type wave function expansions from the DMRG(7,7)[$m$] calculations with
different renormalized active-system states $m$ for the [Fe(NO)]$^{2+}$ molecule surrounded by four point charges
at a distance of $d_{\rm pc} = 0.598$ \AA{} from the iron atom. The CAS(7,7)SCF reference calculation is also shown for comparison.
thr corresponds to the threshold value of COM in the sampling-reconstruction procedure and denotes the accuracy of the
reconstructed CASCI-type wave function. All Slater determinants with CI coefficients in an interval as indicated on the
abscissa are grouped together.
}\label{fig:feno_ci_pattern}
\end{figure}

Similarly to the single-reference problem discussed above, reliable spin densities obtained from reduced
dimensional CASCI-type wave function expansions can be determined for a COM $\geq0.001$ independent of $m$.
A complete collection of spin density distributions for different CASCI-type wave function expansions 
and DMRG parameter sets can be found in the Supporting Information.
Figure \ref{fig:feno_ci_pattern} shows the ratio of Slater determinants with respect to the complete
$N$-particle Hilbert space which have been picked up in the sampling procedure and sorted by their corresponding
CI weights for the DMRG(7,7)[32] and DMRG(7,7)[64] calculation.
For COM $\geq0.001$, the reconstructed CASCI-type wave function contains the major part of the important Slater
determinants, while for a further decreased threshold value of $10^{-5}$
almost all significant Slater determinant have been picked up. Note that the sampling procedure was restricted
to accept only configurations with CI coefficients larger than the threshold value for COM.
Although all possible excitations are included in the CASCI-type wave function in the limit of COM $\rightarrow$ 0
(see also Figure~4 of the Supporting Information), the pattern of the CI coefficients of the DMRG(7,7)[64] calculation
is different from the CI pattern of the CAS(7,7)SCF reference.
While large CI coefficients ($> 0.0001$) are reproduced within sufficient accuracy, smaller CI weights are
underestimated. The maximum of the curve is shifted towards smaller CI weights $<10^{-7}$. Hence, the DMRG algorithm
disregards an exact weighting of unimportant configurations with small
CI coefficients which is a feature of matrix product and tensor network states where large CI coefficients should be reproduced
and unimportant configurations are neglected \cite{cgtnarxiv2010,marti2010pccp}.
A complete collection of excitation histograms for different CASCI-type wave functions can be found in the Supporting Information.
To quantify the differences in the underlying wave functions for our three DMRG calculations employing
$m_i \in \{32,48,64\}$ active-system states, we calculated the quantum fidelity $F_{m_i,m_{i+1}}$ which forms in this case a set of overlap measures of
\{0.831887,0.897445,0.955669\}.

We conclude that reliable spin density distributions can be calculated either from converged DMRG
ground state wave functions or from the reconstructed CASCI-type wave function expansions. In particular,
a fully converged DMRG wave function is not mandatory to obtain qualitatively correct spin density distributions
if the CI weights of the most important configurations are well reproduced for a given $m$-value.
This holds for both the single-reference and the multi-reference case.
A representative set of Slater determinants, i.e., the most important ones ($|C_I|$ $> 0.001$), is sufficient for
a qualitatively correct spin density distribution.

%%%%%%%%%%%%%%%%%%%%%%%%%%%%%%%%%%%%%%%%%%%%%%%%%%%%%%%%%%%%%%%%%%%%%%%%%%%%%%%%%%%%%%%%%%%5
\section{Spin density distributions for large active spaces \label{Sec:results}}
%%%%%%%%%%%%%%%%%%%%%%%%%%%%%%%%%%%%%%%%%%%%%%%%%%%%%%%%%%%%%%%%%%%%%%%%%%%%%%%%%%%%%%%%%%%5
While we have studied the convergence features of DMRG calculations for small active spaces, for which we could 
obtain an exact CASSCF reference result, we shall now proceed to explore territory with DMRG that is not
accessible to the CASSCF approach.
In our recent analysis of CASSCF spin densities for the [Fe(NO)]$^{2+}$ molecule\cite{feno}, the spin density distribution was qualitatively
converged with respect to the dimension of the active orbital space. For quantitatively accurate spin densities we need to 
increase the dimension of the active orbital space so that important iron and ligand orbitals which are missing in the standard CASSCF calculations,
e.g., the Fe $d_{x^2-y^2}$ double-shell orbital, could also be included in the active orbital space.
Here, we extend the convergence series presented in Ref.~\cite{feno} by considering active orbital spaces containing 
up to 29 active orbitals. Starting with an active orbital space comprising 13 active electrons correlated in 20 active orbitals,
the number of active orbitals is further increased to 24 and 29, respectively. The two largest active orbital spaces do also
contain the fifth $d_{x^2-y^2}$-double-shell orbital which could not be included in all CASSCF calculations presented in Ref.~\cite{feno}.
The [Fe(NO)]$^{2+}$ molecular structure features the same bond distances and angles as presented in Section \ref{Sec:test}.
The four point charges of $-$0.5 $e$ are located at a distance of 1.131 \AA{} from the metal center in order to properly
model the square-planar ligand field of the full-fledged complexes in a doublet spin state.

\begin{figure}[h]
\centering
\includegraphics[width=0.9\linewidth]{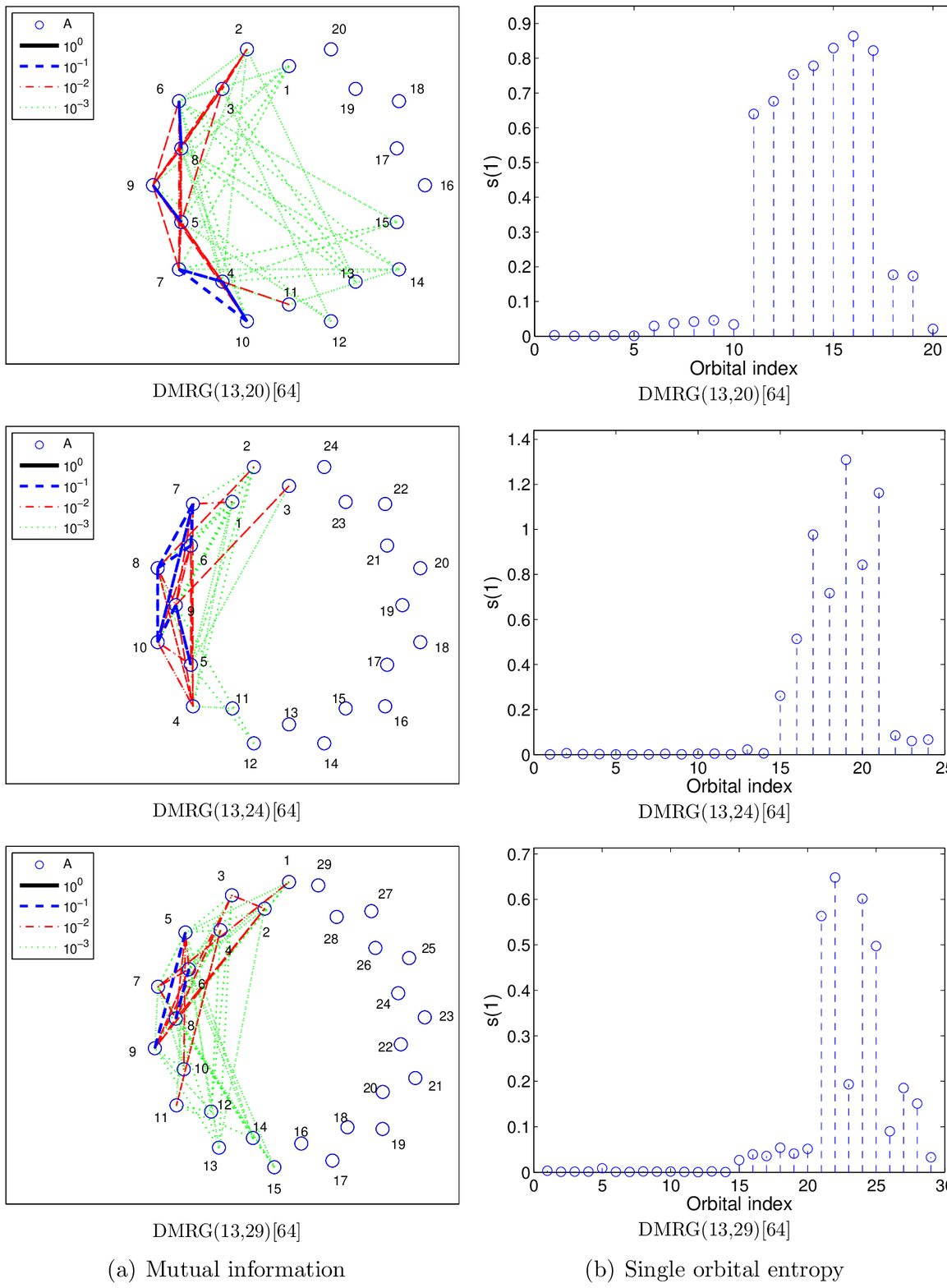}
\caption{Mutual information and single orbital entropies $s(1)$ for the DMRG(13,$y$)[64] calculations determined for different
numbers of active orbitals in the [Fe(NO)]$^{2+}$ molecule surrounded by four point charges at a distance of
$d_{\rm pc} = 1.131$ \AA{} from the iron center.
}\label{fig:orbital_ordering}
\end{figure}

For all DMRG calculations, the natural orbitals from a CAS(11,14)SCF calculation are employed as orbital basis \cite{werner3,werner,werner2}.
Similarly, the CASSCF calculation as well as the calculations of the one-electron and two-electron integrals in the natural orbital basis
were performed with the \textsc{Molpro} program package \cite{molpro} using Dunning's cc-pVTZ basis set
for all atoms \cite{dunning,dunning2}, while the DMRG calculations are performed with the Budapest DMRG program \cite{dmrg_ors}.
In addition, the DMRG orbital orderings were optimized for all three active orbital spaces and the CI-DEAS starting guess
was performed. Figure~\ref{fig:orbital_ordering} displays the corresponding single orbital entropies given by
\begin{equation}\label{Eq:sd_firstquant}
s(1)_i = - \sum_\alpha \omega_{\alpha,i} \ln \omega_{\alpha,i},
\end{equation}
and mutual information determined by
\begin{equation}\label{Eq:sd_firstquant}
I_{i,j} = s(2)_{i,j} - s(1)_{i} - s(1)_{j},
\end{equation}
where $i = {1\ldots k}$ is the orbital index and runs over all $k$ one-particle states, $\omega_{\alpha,i}$ is the $\alpha$ eigenvalue of the reduced density matrix
of orbital $i$ \cite{legeza_dbss}, while $s(2)_{i,j}$ is the two-orbital entropy between a pair ${i,j}$ of sites introduced by Rissler \emph{et al.}
to the quantum chemical DMRG algorithm \cite{Rissler2006519}.
Note that the mutual information and single orbital entropies are confined to the first ten natural orbitals for all considered dimensions
of the active orbital space. These natural orbitals are highly entangled and represent the most important orbitals comprised in the active
orbital space. Therefore, accurate DMRG spin densities can be obtained already for a reasonably small number of active-system states.
Similar entropy profiles can be obtained for smaller dimensions of the active orbital space.

The number of DMRG active-system states $m$ was set to 128, 256, 512, 1024 and 2048, respectively.
The ground state energies for all DMRG calculations are summarized in Table \ref{tab:energieslarge}. Considering the DMRG(13,20)[$m$]
calculations, an energy convergence of 0.135 mH (0.4 kJ/mol) is reached with respect to $m$. For the largest active orbital space,
the DMRG(13,29)[1024] energy is converged to 1.195 mH (3.1 kJ/mol) when compared to the DMRG(13,29)[2048] reference.
\begin{table}[h]
\caption{Ground state energy for [Fe(NO)]$^{2+}$ surrounded by four point charges at a distance of
$d_{\rm pc} = 1.131$ \AA{} from the iron center in Hartree atomic units for our DMRG($x$,$y$)[$m$] calculations employing
different numbers of DMRG active-system states $m$. The CAS(11,14)SCF energy is $-$1393.013 396 Hartree.}\label{tab:energieslarge}
{
\begin{center}
\begin{tabular}{lccc}\hline \hline
& \multicolumn{3}{c}{$E$/Hartree} \\\hline
Method & DMRG(13,20) & DMRG(13,24) & DMRG(13,29) \\\hline
$m=128$ &  $-$1393.014 662 & $-$1392.991 085 & $-$1393.014 010 \\
$m=256$ &  $-$1393.018 626 & $-$1393.019 309 & $-$1393.024 883 \\
$m=512$ &  $-$1393.020 065 & $-$1393.021 876 & $-$1393.030 374 \\
$m=1024$ & $-$1393.020 511 & $-$1393.022 946 & $-$1393.033 001 \\ 
$m=2048$ & $-$1393.020 646 & $-$1393.023 294 & $-$1393.034 196 
\\ \hline
\hline
\end{tabular}
\end{center}
}
\end{table}

%%%%%%%%%%%%%%%%%%%%%%%%%%%%%%%%%%%%%%%%%%%%%%%%%%%%%%%%%%%%%%%%%%%%%%%%%%%%%%%%%%%%%%%%%%%5
\subsection{Convergence of DMRG spin densities}
%%%%%%%%%%%%%%%%%%%%%%%%%%%%%%%%%%%%%%%%%%%%%%%%%%%%%%%%%%%%%%%%%%%%%%%%%%%%%%%%%%%%%%%%%%%5

\begin{figure}[h]
\centering
\includegraphics[width=0.8\linewidth]{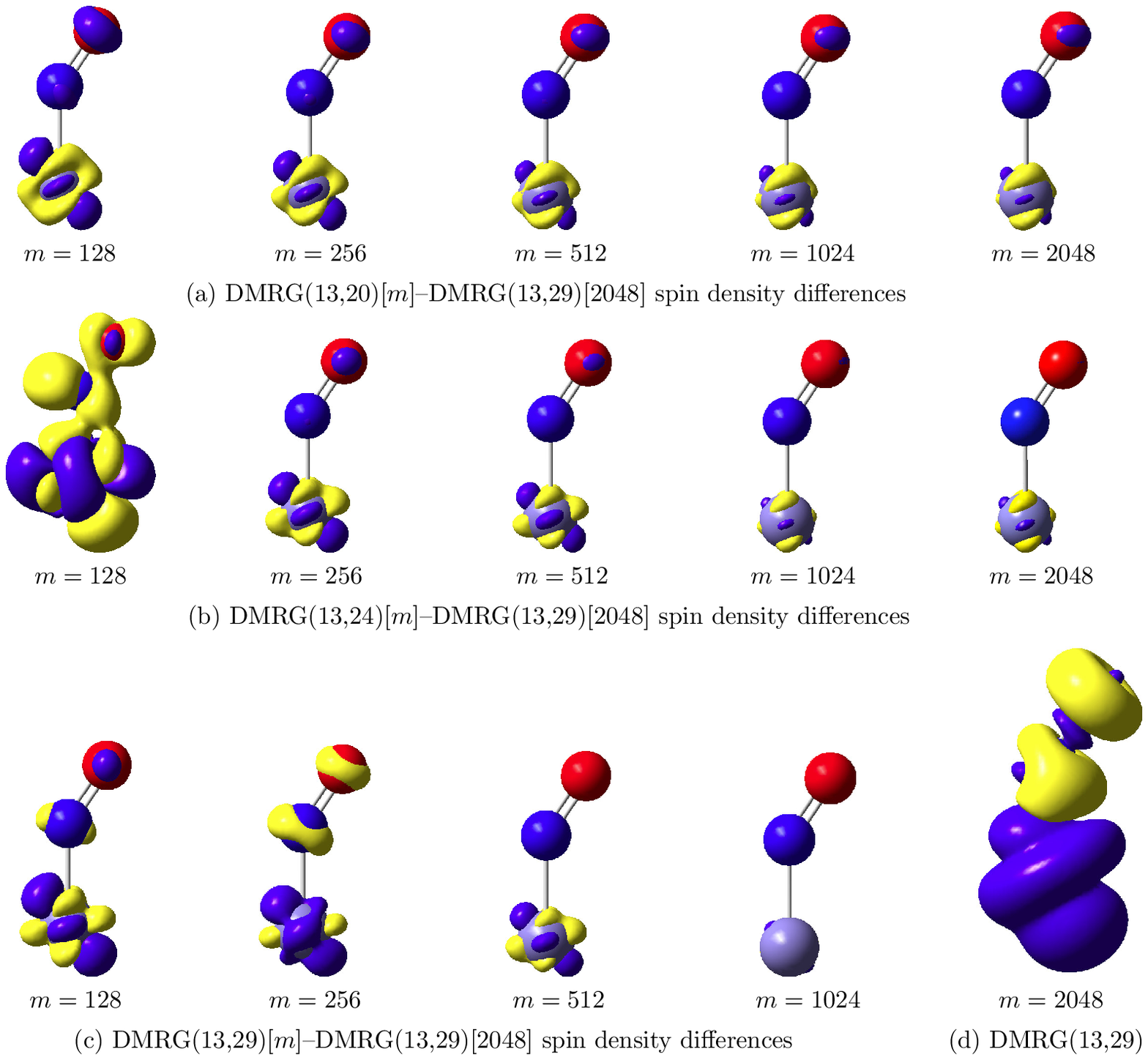}
\caption{DMRG(13,$y$)[$m$] and CAS($x$,$y$)SCF spin density difference plots with respect to the DMRG(13,29)[2048] spin density
distribution (d) for [FeNO]$^{2+}$ surrounded by four point charges at a distance of $d_{\rm pc} = 1.131$ \AA{} from
the iron center. All spin densities are displayed for an isosurface value of 0.001. (a) DMRG(13,20)[$m$]--DMRG(13,29)[2048] spin density difference plots.
(b) DMRG(13,24)[$m$]--DMRG(13,29)[2048] spin density difference plots.
(c) DMRG(13,29)[$m$]--DMRG(13,29)[2048] spin density difference plots. (d) The DMRG(13,29)[2048] reference
spin density distribution.
}\label{fig:large_sd_dmrg}
\end{figure}

The dependence of the spin density distribution on the number of DMRG active-system states $m$ is shown in Figure~\ref{fig:large_sd_dmrg}
where the differences in spin density distribution are plotted for DMRG(13,$y$)[$m$] calculations with
respect to the converged DMRG(13,29)[2048] reference calculation. For increasing $m$-values, the
differences in the spin density distribution decrease (see each row in Figure~\ref{fig:large_sd_dmrg} from the left to the right).
Similarly, we observe
that the spin density gradually converges with respect to the dimension of the active orbital space (see last
column from the top to the bottom of Figure~\ref{fig:large_sd_dmrg}). In particular, changes in the spin density are 
negligible when $m$ is increased from 1024 to 2048, and hence, reliable spin density distributions can be obtained
even if the total energy is not yet converged with respect to $m$ (the difference is 1.195 mH, see above).

Furthermore, the $\Delta_{\rm abs}$- and $\Delta_{\rm rs}$-values quantify the convergence series of the determined
DMRG spin density distributions. In Table \ref{tab:error_dens}, both error quantities are listed for
each DMRG($x$,$y$)[$m$] spin density with respect to the DMRG(13,29)[2048] reference spin density. In general, the absolute error
$\Delta_{\rm abs}$ and the root-square error $\Delta_{\rm rs}$ decrease for increasing $m$ keeping
the dimension of the active orbital space fixed.
Note that larger active orbital spaces require a larger $m$-value to obtain
the same accuracy as achieved in smaller active space calculations. This is not immediately evident from the
error data presented in Table \ref{tab:error_dens} since different dimensions of the active orbital space are
compared which result in nonzero error values, while error values determined for different parameter sets, but
the same dimension of the active orbital space could vanish. The large error values for the DMRG(13,24)[128] calculation
indicate that important states were not picked up by the DMRG algorithm resulting in the large differences in
the spin density distribution displayed in Figure~\ref{fig:large_sd_dmrg}. Furthermore, since both error values determined for
the DMRG(13,29)[1024] calculation are below the threshold values, no considerable improvement in the accuracy of
the spin density distribution can be expected when $m$ is further increased to more than 2048 active-system states.

\begin{table}[h]
\caption{The absolute error $\Delta_{\rm abs}$ and the root-square error $\Delta_{\rm rs}$ of the DMRG(13,$y$)[$m$] spin
densities with respect to the converged DMRG(13,29)[2048] reference spin density for a different number of normalized
active-system states $m$ for [FeNO]$^{2+}$ surrounded by four point charges at a distance of $d_{\rm pc}=1.131$ \AA{} from
the iron center. The $\Delta_{\rm abs}$ and $\Delta_{\rm rs}$ values of the CAS($x$,$y$)SCF calculations
of Ref.~\cite{feno} with respect to the DMRG(13,29)[2048] reference spin density are also listed.
}\label{tab:error_dens}
{\small
\begin{center}
\begin{tabular}{lcc}\hline \hline
Method           &  $\Delta_{\rm abs}$ & $\Delta_{\rm rs}$ \\ \hline
DMRG(13,20)[128] &  0.030642  & 0.008660 \\
DMRG(13,20)[256] &  0.020088  & 0.004930 \\
DMRG(13,20)[512] &  0.016415  & 0.003564 \\
DMRG(13,20)[1024]&  0.015028  & 0.003162 \\
DMRG(13,20)[2048]&  0.014528  & 0.003028 \\ \hline
DMRG(13,24)[128] & 0.590022 & 0.235922 \\
DMRG(13,24)[256] & 0.020993 & 0.003245 \\
DMRG(13,24)[512] & 0.014045 & 0.003633 \\
DMRG(13,24)[1024]& 0.011622 & 0.002668 \\
DMRG(13,24)[2048]& 0.010731 & 0.002361 \\\hline
DMRG(13,29)[128] & 0.032171 & 0.010677 \\
DMRG(13,29)[256] & 0.026005 & 0.006790 \\
DMRG(13,29)[512] & 0.010826 & 0.003406 \\        
DMRG(13,29)[1024]& 0.003381 & 0.000975 \\ \hline
CAS(11,11)SCF    & 0.086658 & 0.024495 \\
CAS(11,12)SCF    & 0.080249 & 0.020591 \\ 
CAS(11,13)SCF    & 0.046303 & 0.011402 \\
CAS(11,14)SCF    & 0.042544 & 0.010954 \\
CAS(13,13)SCF    & 0.052239 & 0.012124 \\
CAS(13,14)SCF    & 0.073400 & 0.019850 \\
CAS(13,15)SCF    & 0.053157 & 0.011180 \\
CAS(13,16)SCF    & 0.104928 & 0.031922
\\ \hline
\hline
\end{tabular}
\end{center}
}
\end{table}

In order to demonstrate the convergence of the DMRG(13,29) wave function with respect to $m$ (and thus the convergence of the
obtained DMRG(13,29)[2048] reference spin density distribution),
the CASCI-type wave function expansions are reconstructed and compared for all $m$-values.
In particular, the influence of the missing $d_{x^2-y^2}$-double-shell orbital can be assessed by examining the CI coefficients
corresponding to Slater determinants with an occupied $d_{x^2-y^2}$-double-shell orbital. 
Following the conclusions of a benchmark study for intermediate CAS sizes (see Supporting Information), only the most important configurations ($|C_I|$ $\geq$ 0.00001) are necessary
to obtain an accurate wave function expansion. As convergence threshold for the sampling procedure,
a value of 0.001 is sufficient. 
With this threshold, similar CASCI-type wave function expansions are obtained for a
quantum fidelity measure close to 0.998.
The set of quantum fidelity measures $F_{m_i,m_{i+1}}$ for our five DMRG
calculations with $m_i \in \{128,256,512,1024,2048\}$ is \{0.991800,0.995510,0.996983,0.997639\}.
As the number of DMRG active-system states is enlarged, the CI coefficients of the
reconstructed wave function expansion converge gradually which is indicated by the increasing quantum fidelity
measure. Note that $F_{m_1,m_2}$ is close to the ideal value of 0.998 already for a small number of DMRG
active-system states $m$, and hence, only minor variations in the large CI coefficients occur when $m$ is increased
which explains the slight differences in the spin density distributions displayed in
Figure~\ref{fig:large_sd_dmrg}(c). 

\begin{figure}[h]
\centering
\includegraphics[width=0.8\linewidth]{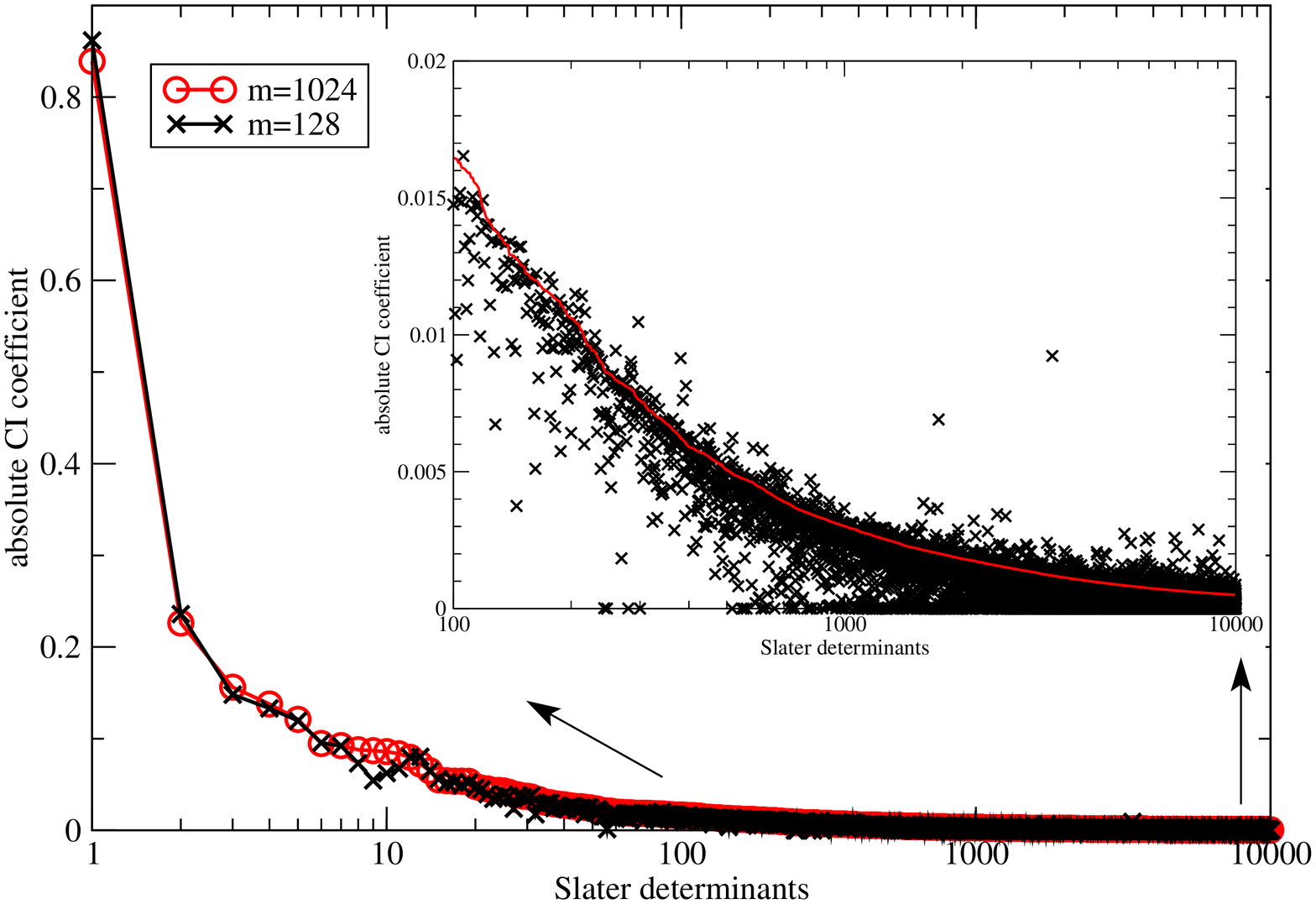}
\caption{Distribution of the absolute value of the CI coefficients for the DMRG(13,29)[$m$] calculations with
$m=128$ and 1024, respectively, for [FeNO]$^{2+}$ surrounded by four point charges at a distance of $d_{\rm pc} =1.131$ \AA{} from
the iron center. The CI coefficients reconstructed for both DMRG calculations are always
printed for the same Slater determinants. The determinants are ordered according to the CI weight of the
DMRG(13,29)[2048] reference calculation.
}\label{fig:large_128-1024}
\end{figure}

To demonstrate that this is indeed the case, the CI coefficients of the most important Slater
determinants ($|C_I|$ $> 0.0001$) corresponding to the $m=128$ and $m=1024$ calculations are shown in Figure~\ref{fig:large_128-1024}.
Slater determinants with large CI weights ($|C_I|$ $> 0.05$) are similar for both DMRG parameter sets, only minor deviations can be
observed. Note that all these Slater determinants have been incorporated in the DMRG wave function already for $m=128$.
Considerable differences in CI weights are present for Slater determinants corresponding to small-valued
CI coefficients ($|C_I|$ $< 0.015$), while some Slater determinants with $|C_I| < 0.01$ have not been
incorporated in the DMRG wave function for $m=128$. These off-size or missing configurations lead to the different
spin density distributions for small-$m$ values.

From the reconstructed CASCI-type wave function, the influence of the $d_{x^2-y^2}$-double-shell orbital
as well as of the empty ligand orbitals on the spin density distribution can be analyzed. In the upper part of Table \ref{tab:ci_large},
configurations containing an occupied
$d_{x^2-y^2}$-double-shell orbital and corresponding to the largest CI coefficients are presented. In the lower part
of Table~\ref{tab:ci_large}, some selected
configurations with large CI coefficients carrying excitations to empty ligand orbitals that can not be included in standard CASSCF
calculations are presented and compared for the DMRG(13,29)[128] and DMRG(13,29)[1024] calculations. In general, Slater determinants with
an occupied $d_{x^2-y^2}$-double-shell orbital feature small CI weights ($\leq 0.003$) and are hence of minor importance, while Slater determinants bearing occupied ligand orbitals
feature large CI coefficients. Configurations containing occupied ligand orbitals that are only included in
the DMRG(13,29)[$m$] calculations (marked in bold face in Table~\ref{tab:ci_large}) possess considerably large CI weights. All other Slater determinants with
excitations to different empty ligand orbitals have smaller CI coefficients. 
Hence, those ligand orbitals pose a significant contribution in obtaining accurate spin density distributions for the small model complex
and cannot be neglected from the active orbital space.

\begin{table}[h]
\caption{Some important occupation number vectors (ONV) with the corresponding CI weights from DMRG(13,29)[$m$] calculations for [FeNO]$^{2+}$
surrounded by four point charges at a distance of $d_{\rm pc} = 1.131$ \AA{} from
the iron center. Upper part: ONVs containing an occupied $d_{x^2-y^2}$-double-shell orbital (marked in bold face).
Bottom part: additional selected important configurations with occupied natural orbitals that cannot be included in the active orbital space
in CASSCF calculations (marked in bold face), for the same DMRG(13,29)[$m$] calculations.
2: doubly occupied natural orbital; a: natural orbital occupied by an $\alpha$-electron; b:
natural orbital occupied by a $\beta$-electron; 0: empty natural orbital.
}\label{tab:ci_large}
{
\begin{center}
\begin{tabular}{crr}\hline \hline
     & \multicolumn{2}{c}{CI weight} \\ \cline{2-3}
Slater determinant     & $m=128$ & $m=1024$ \\\hline
b2b222a0a0000000  0000000 {\bf a} 00000 &    0.003 252 &    0.003 991 \\	
bb2222aa00000000  0000000 {\bf a} 00000 & $-$0.003 226 & $-$0.003 611 \\
222220ab00000000  0000000 {\bf a} 00000 & $-$0.002 762 & $-$0.003 328 \\
ba2222ab00000000  0000000 {\bf a} 00000 &    0.002 573 &    0.003 022 \\
b2a222a0b0000000  0000000 {\bf a} 00000 & $-$0.002 487 & $-$0.003 017 \\
202222ab00000000  0000000 {\bf a} 00000 &    0.002 405 &    0.002 716 \\ \hline
b222a2a0b0000000  {\bf 0000000 0 0000a} &    0.010 360 &    0.011 558 \\
22b2a2a0a0000000  {\bf 0000000 0 b0000} &    0.009 849 &    0.011 366 \\
22b2a2a0b0000000  {\bf 0000000 0 a0000} & $-$0.009 532 & $-$0.011 457 \\
b2222aab00000000  {\bf 0000000 0 0000a} & $-$0.009 490 & $-$0.010 991 \\
a2222baa00000000  {\bf 0000000 0 0000b} & $-$0.009 014 & $-$0.010 017 \\
b2b222a0a0000000  {\bf 0000000 0 0a000} &    0.008 820 &    0.010 327 \\
b2222aab00000000  {\bf 00a0000 0 00000} & $-$0.004 277 & $-$0.005 436 \\
22b2a2a0b0000000  {\bf a000000 0 00000} & $-$0.004 224 & $-$0.006 852 
\\ \hline
\hline
\end{tabular}
\end{center}
}
\end{table}

%%%%%%%%%%%%%%%%%%%%%%%%%%%%%%%%%%%%%%%%%%%%%%%%%%%%%%%%%%%%%%%%%%%%%%%%%%%%%%%%%%%%%%%%%%%5
\subsection{Assessment of CASSCF spin densities}
\begin{figure}[h]
\centering
\includegraphics[width=0.7\linewidth]{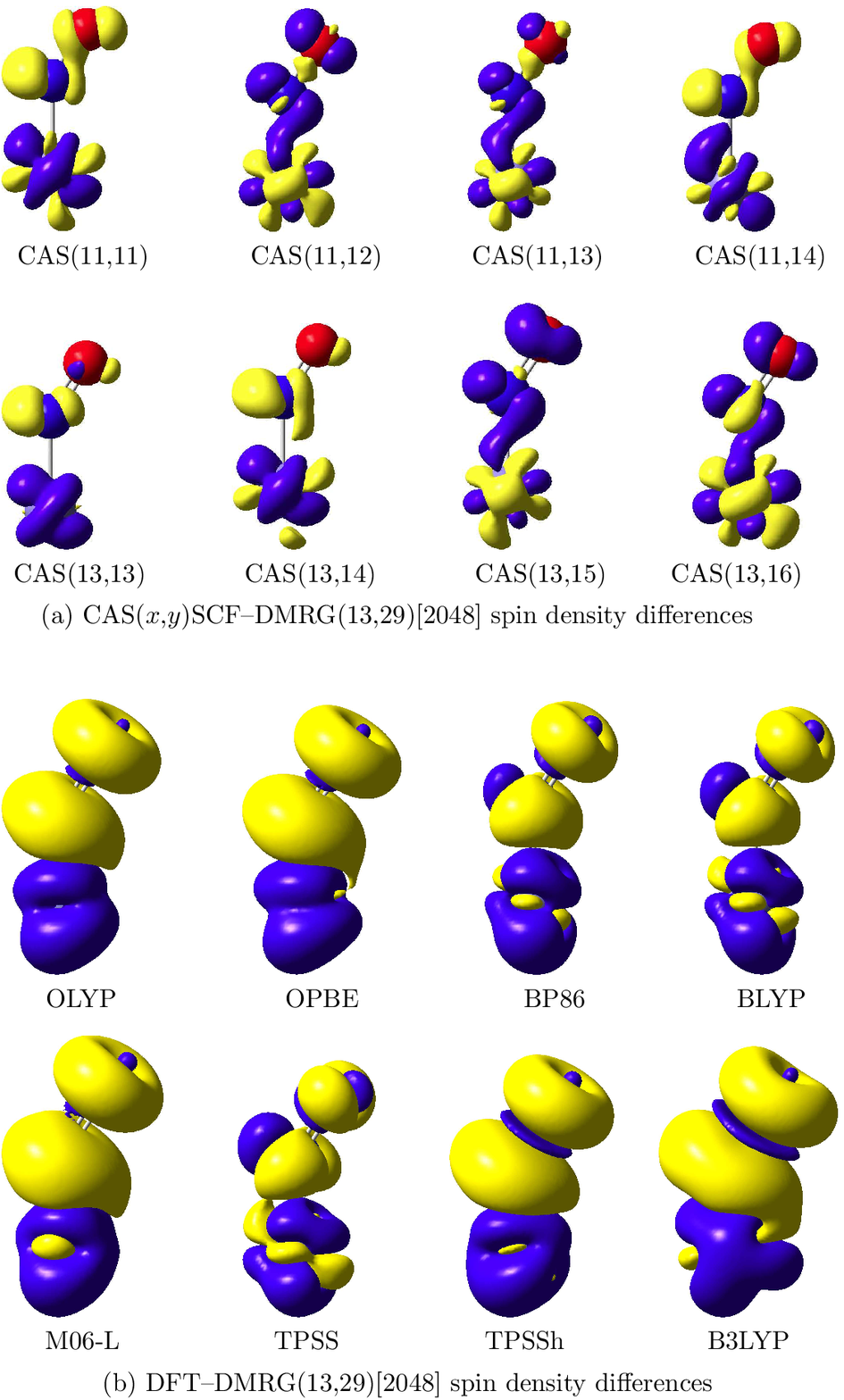}
\caption{(a) CAS($x$,$y$)SCF and (b) DFT spin density difference plots with respect to the DMRG(13,29)[2048] spin density
distribution for [FeNO]$^{2+}$ surrounded by four point charges at a distance of $d_{\rm pc} = 1.131$ \AA{} from
the iron center. All spin densities are displayed for an isosurface value of 0.001.
}\label{fig:large_sd_casscf}
\end{figure}

%%%%%%%%%%%%%%%%%%%%%%%%%%%%%%%%%%%%%%%%%%%%%%%%%%%%%%%%%%%%%%%%%%%%%%%%%%%%%%%%%%%%%%%%%%%5
The converged DMRG(13,29)[2048] reference spin density can be used to assess the accuracy of CASSCF
spin density distributions and benchmark the quality of the (restricted) active orbital spaces
in standard CASSCF calculations (see Figure~\ref{fig:large_sd_casscf}(a)).
Note that the same isosurface value has been taken to display the DMRG(13,$y$)[$m$]--DMRG(13,29)[2048]
and CAS($x$,$y$)SCF--DMRG(13,29)[2048] spin density difference plots.
The CASSCF spin density distributions determined for medium-sized active orbital spaces oscillate around the 
converged DMRG spin density. 
Depending on which double-$d$-shell orbital is included in the active orbital space,
the $\beta$-electron density around the NO ligand is either overestimated or underestimated. This results either in pure
spin-polarized cases with $\beta$-electron density found only around the nitrosyl ligand for CAS(11,11), CAS(11,14), CAS(13,13), and CAS(13,14),
or some additional $\alpha$-electron density present around the NO ligand associated with a simultaneous
decrease in the $\beta$-electron density for
CAS(11,12), CAS(11,13), CAS(13,15), and CAS(13,16).

Similarly, the large $\Delta_{\rm abs}$- and $\Delta_{\rm rs}$-values stress the differences in the spin density
distributions which are considerably larger than those from the DMRG(13,$y$)[$m$]--DMRG(13,29)[2048]
difference analysis (Table \ref{tab:error_dens}).
Furthermore, Table \ref{tab:error_dens} indicates that the CAS(11,11)SCF- and CAS(11,12)SCF calculations and
the CAS(11,13)SCF- and CAS(11,14)SCF calculations, respectively, are of similar accuracy, as they have similar error values,
but the spin density difference plots emphasize the $qualitatively$ different spin density distributions.
Increasing the dimension of the active orbital space results in even larger deviations from the DMRG reference spin density
because the active space is not stable and important orbitals are rotated out of the CAS.
Note that all DMRG calculations---except DMRG(13,24)[128]---yield smaller error-values and smaller differences
in the spin density difference plots.

Although the CASSCF spin densities are quantitatively converged with respect to the active orbital space,
significant qualitative---but also non-negligible quantitative---differences to the DMRG(13,29)[2048]
reference spin density can be observed.
The extension of the active orbital space by including an additional shell of $d$-orbitals only is not sufficient to obtain a
$qualitatively$ accurate spin density distribution for the small iron nitrosyl molecule. Our analysis indicates that empty ligand orbitals
are essential for calculating reliable reference spin densities.
This may have severe implications for the standard CASSCF approach that require further analysis in future work.

%%%%%%%%%%%%%%%%%%%%%%%%%%%%%%%%%%%%%%%%%%%%%%%%%%%%%%%%%%%%%%%%%%%%%%%%%%%%%%%%%%%%%%%%%%%5
\subsection{Comparison to DFT spin densities}
%%%%%%%%%%%%%%%%%%%%%%%%%%%%%%%%%%%%%%%%%%%%%%%%%%%%%%%%%%%%%%%%%%%%%%%%%%%%%%%%%%%%%%%%%%%5
A comparison of DFT and CASSCF spin density distributions for medium-sized active orbital spaces for
the [FeNO]$^{2+}$ molecule has been already discussed in our previous work (see Ref.~\cite{feno} for
more details). For an unambiguous benchmark of approximate exchange--correlation density functionals, the DFT
spin densities of Ref.~\cite{feno} can be compared to the DMRG reference distribution.
The qualitative analysis of the DFT--DMRG(13,29)[2048] spin density difference distributions is shown
in Figure~\ref{fig:large_sd_casscf}(b).
When comparing to the results obtained in Ref.~\cite{feno}, similar conclusions concerning the
performance of approximate exchange--correlation density functionals can be drawn. The best agreement is found for BP86, BLYP, and TPSS, while the
remaining approximate exchange--correlation density functionals yield larger deviations and result in too large
spin polarization. We should note that BP86, BLYP, and TPSS correctly predict the distribution of the
$\alpha$-electron density around the nitrosyl ligand, although it is overemphasized. In general, nonhybrid functionals yield spin
densities which are in closest agreement with the DMRG reference distributions. This observation is
supported by both error measures which are smallest for BP86, BLYP, and TPSS (see Table I in the Supporting
Information).

%%%%%%%%%%%%%%%%%%%%%%%%%%%%%%%%%%%%%%%%%%%%%%%%%%%%%%%%%%%%%%%%%%%%%%%%%%%%%%%%%%%%%%%%%%%5
\section{Conclusions and Outlook\label{Sec:conclusion}}
%%%%%%%%%%%%%%%%%%%%%%%%%%%%%%%%%%%%%%%%%%%%%%%%%%%%%%%%%%%%%%%%%%%%%%%%%%%%%%%%%%%%%%%%%%%5
In this work, we have demonstrated how reliable \emph{ab initio} spin density distributions can be calculated
for very large active spaces. Our procedure is based on the DMRG algorithm and on two
different approaches to obtain spin density matrix elements: (i)
on-the-fly directly from the second-quantized DMRG elementary operators or
(ii) from an approximate CASCI-type wave function expansion which is determined by our
SRCAS algorithm \cite{srcas_jcp}. The reconstructed CASCI-type wave function can also be used as
a means to compare a series of DMRG calculations employing a different number
of DMRG active-system states $m$.

The small noninnocent molecule [FeNO]$^{2+}$ surrounded by four point charges
represents a suitable system to validate our approach.
The spin density distributions are highly sensitive to the nature of the converged state.
We deliberately converged DMRG wave functions that correspond to local minima in the
electronic energy in order to compare with qualitatively wrong wave functions.
The possibility of convergence into local minima is shown by examining the (largest) CI coefficients
of the SRCAS-reconstructed CASCI-type wave function.
Strong deviations with respect to the absolute value of the CI coefficients indicate that the number
of DMRG active-system states $m$ is chosen too small, and hence important states have not been incorporated
by the DMRG algorithm.
Spin densities corresponding to such local minima deviate considerably from the ground state spin density.

The convergence analysis of the spin density distribution for the [FeNO]$^{2+}$ molecule considered
active orbital spaces comprising up to 29 active orbitals. Difference plots of the spin
density distribution for different active orbital spaces as well as the absolute error and the root-square 
error in the spin density difference distribution indicate a $quantitatively$ converged spin density
with respect to the dimension of the active orbital space and the number of active-system states $m$ (which was as large as
$m=2048$). The DMRG reference spin density has been used to validate CASSCF spin densities resulting in significant
$quantitative$ and even $qualitative$ differences.
Considering an additional shell of $d$-orbitals is not sufficient to obtain reliable spin densities
for the small model system and the active orbital space must be extended by additional unoccupied ligand orbitals.
Similar difficulties are likely to be present for larger iron nitrosyl complexes where the point charges are replaced by different ligands
and hence additional ligand and iron orbitals must be included in the active orbital space.
The DMRG study of larger \{FeNO\}$^7$ complexes is now pursued in our laboratory.

A convergence analysis of the spin density in terms of spin density difference plots
with respect to the number of DMRG active-system states indicates that reliable reference spin densities can be
obtained even if total energies are not converged with respect to $m$. A similar conclusion was found in
our previous work regarding the energy splittings of states of different spin multiplicity
\cite{marti2008,marti2010b,orbitalordering}.
Comparison of CI weights corresponding to the most
important configurations of the reconstructed CASCI-type wave functions for different $m$-values
furthermore ensures that reliable spin densities are obtained.
The similarities in DMRG wave functions can be quantified by the quantum fidelity measure
which can be used as an additional convergence criterion for spin density distributions
in a sequence of DMRG calculations.

Spin densities calculated from approximate CASCI-type wave functions are in good agreement with the
DMRG reference spin density. Qualitatively reliable spin densities can be obtained even for large thresholds of COM
($0.001$) when the most important configurations have been picked up in the wave function expansion.
For this threshold, the CASCI-type wave function contains Slater determinants with CI weights larger than
$0.00001$ which are important for the spin density.

The comparison of DFT spin densities with the DMRG reference distributions allows us to benchmark
approximate exchange--correlation density functionals.
Although nonhybrid functionals yield spin density distributions closest to the DMRG reference,
significant qualitative and quantitative differences to the DMRG reference distributions could be observed for
all investigated density functionals. Similar conclusions were drawn in our previous study,
where DFT spin densities were assessed against CASSCF spin densities \cite{feno},
entailing that none of the investigated exchange--correlation density functionals yields
sufficiently accurate spin density distributions for the [FeNO]$^{2+}$ molecule.

\section*{Acknowledgments}

We gratefully acknowledge financial support by a TH-Grant (TH-26 07-3) from ETH Zurich, by a grant from
the Swiss national science foundation SNF (project 200020-132542/1),
and from the Hungarian Research Fund (OTKA) under Grant No.~K73455 and K100908.
K.B.\ thanks the Fonds der Chemischen Industrie for a Chemiefonds scholarship.
{\"O}.L.\ acknowledges support from the Alexander von Humboldt foundation and from ETH Zurich
during his time as a visiting professor.

%%%%%%%%%%%%%%%%%%%%%%%%%%%%%%%%%%%%%%%%%%%%%%%%%%%%%%%%%%%%%%%%%%%%%%%%%%%%%%5
%%%%%%%%%%%%%%%%%%%%%%%%%%%%%%%%%%%%%%%%%%%%%%%%%%%%%%%%%%%%%%%%%%%%%%%%%%%%%%5
\section*{Supporting Information}
%%%%%%%%%%%%%%%%%%%%%%%%%%%%%%%%%%%%%%%%%%%%%%%%%%%%%%%%%%%%%%%%%%%%%%%%%%%%%%5
%%%%%%%%%%%%%%%%%%%%%%%%%%%%%%%%%%%%%%%%%%%%%%%%%%%%%%%%%%%%%%%%%%%%%%%%%%%%%%5
Additional details distributions of CI coefficients, excitation patterns and
spin density distributions are available and have been included
in the Supporting Information.
This information is available free of charge via the Internet at http://pubs.acs.org/.

%\bibliographystyle{achemso}
%\bibliography{begin,literatur,end} 

\providecommand{\url}[1]{\texttt{#1}}
\providecommand{\refin}[1]{\\ \textbf{Referenced in:} #1}

\newpage
\begin{center}
{\bf \Large Supporting Information}
\end{center}

\newpage
%\listoffigures
\newpage
\begin{figure}[ht]
\centering
\includegraphics[width=1.0\linewidth]{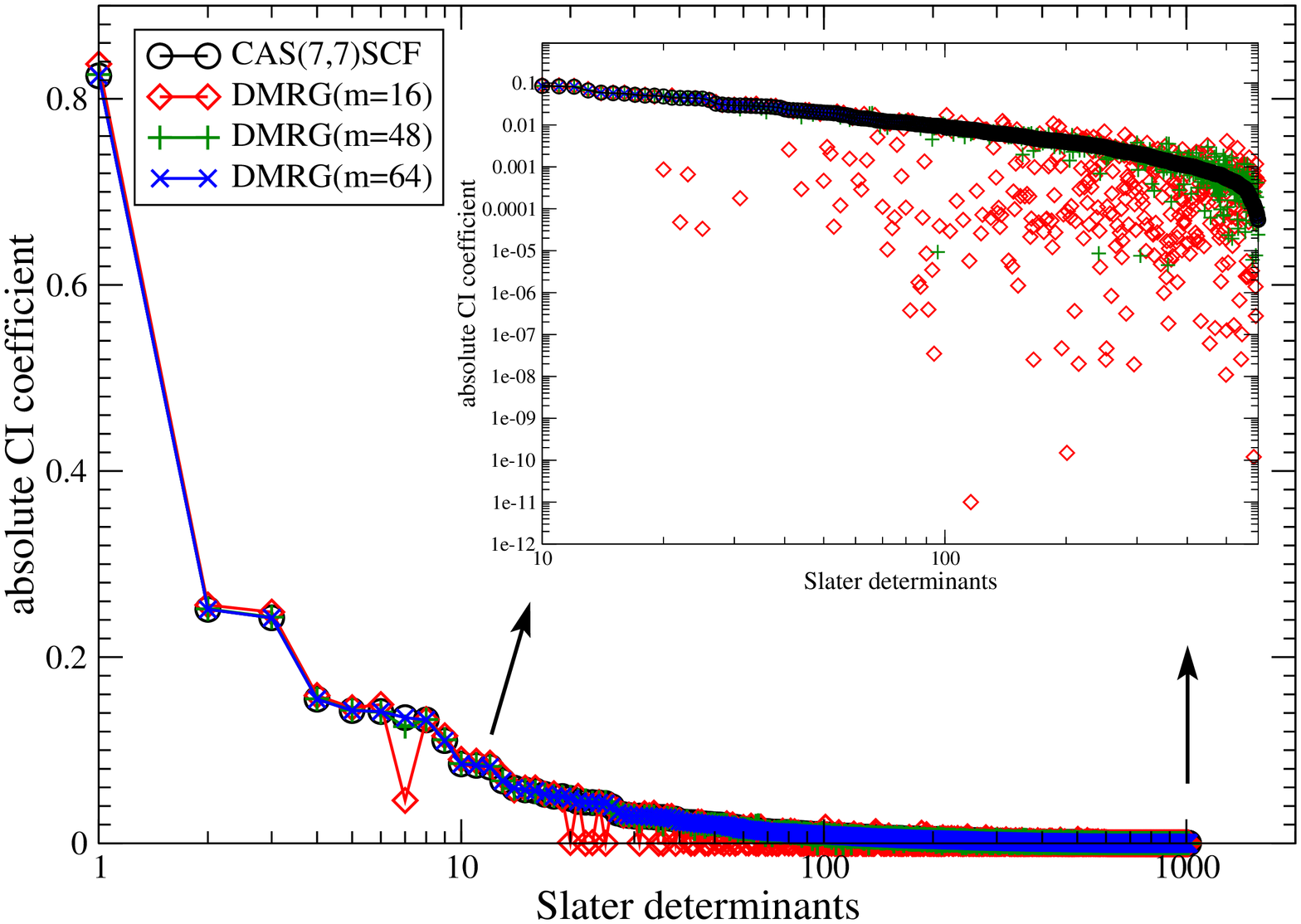}
\caption{Distribution of the absolute value of the CI coefficients for the DMRG(7,7)[$m$] calculations employing different
renormalized active-system states $m$ and the CAS(7,7)SCF reference calculation for the [Fe(NO)]$^{2+}$ molecule surrounded by four
point charges located at a distance of 1.131 \AA{} from the iron center.
The CI coefficient obtained in the different calculations are always printed
for the same Slater determinants. The determinants are ordered according to the CI weights of the CAS(7,7)SCF calculation.}
\end{figure}
\newpage
\clearpage

\begin{figure}[ht]
\centering
\includegraphics[width=0.6\linewidth]{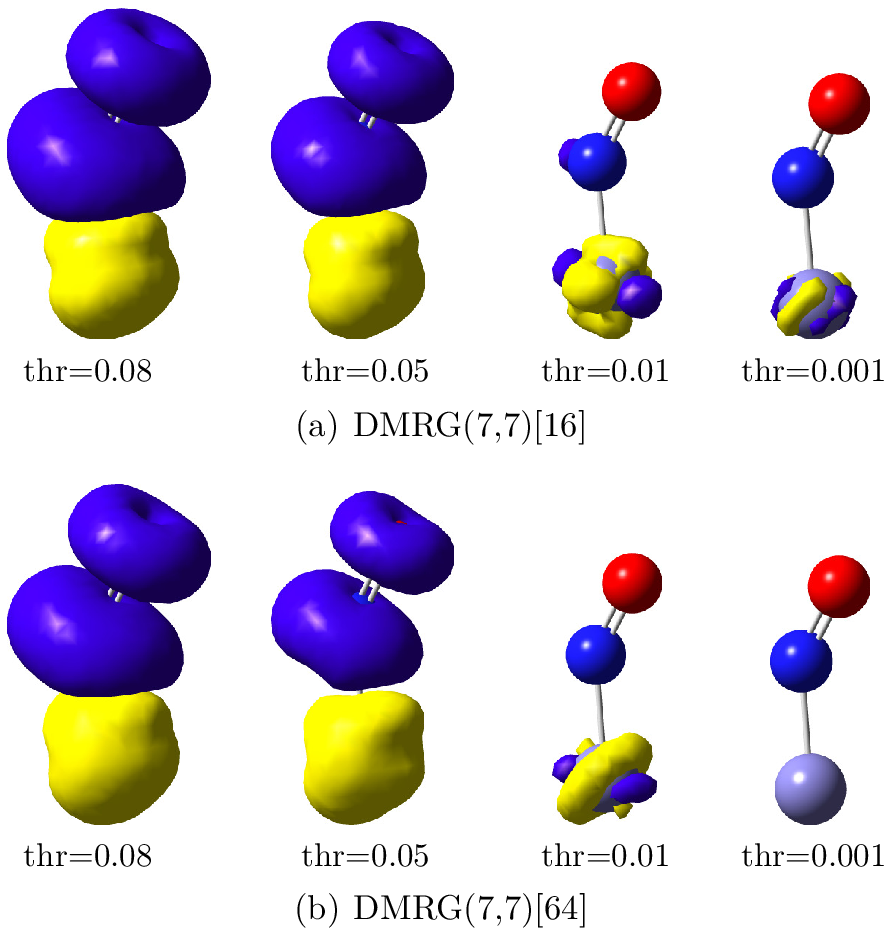}
\caption{CAS(7,7)CI[$m$]--DMRG(7,7)[$m$] spin density difference plots determined for different sampled subspaces of the
complete $N$-particle Hilbert space for the [Fe(NO)]$^{2+}$ molecule surrounded by four
point charges located at a distance of 1.131 \AA{} from the iron center. The spin density differences are plotted
for the reconstructed CAS(7,7)CI-type wave function with respect to the DMRG(7,7)[$m$] reference spin density for the
same number of renormalized active-system states $m$. The threshold (thr) indicates the upper bound of the COM value in
the sampling-reconstruction procedure for each CASCI-type wave function. An isosurface value of 0.0003 was chosen
for all spin density difference plots.
}
\end{figure}
\clearpage

\begin{figure}[ht]
\centering
\includegraphics[width=1.0\linewidth]{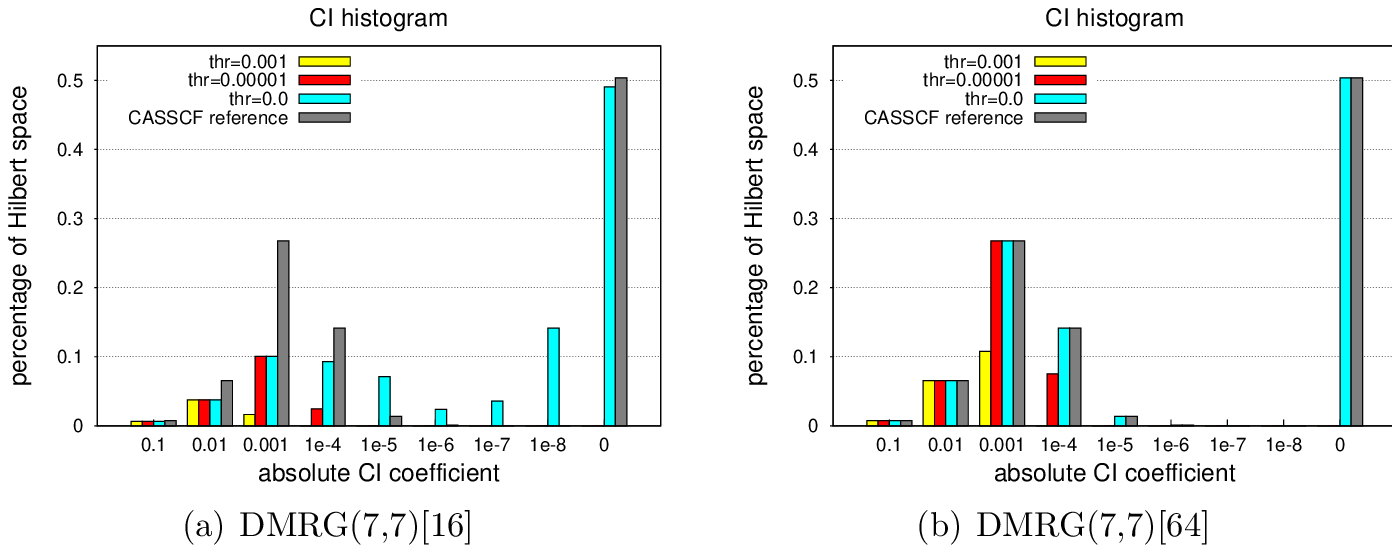}
\caption{CI histogram of the absolute values of the CI coefficients for the Slater determinants
for reconstructed CASCI-type wave function expansions from the DMRG(7,7)[$m$] calculations with
different renormalized active-system states $m$ for the [Fe(NO)]$^{2+}$ molecule surrounded by four point charges
at a distance of $d_{\rm pc} = 1.131$ \AA{} from the iron atom. The CAS(7,7)SCF reference calculation is also shown for comparison.
thr corresponds to the threshold value of COM in the sampling-reconstruction procedure and denotes the accuracy of the
reconstructed CASCI-type wave function. All Slater determinants with CI coefficients in an interval as indicated on the
abscissa are grouped together.
}
\end{figure}
\clearpage

\begin{figure}[ht]
\centering
\includegraphics[width=1.0\linewidth]{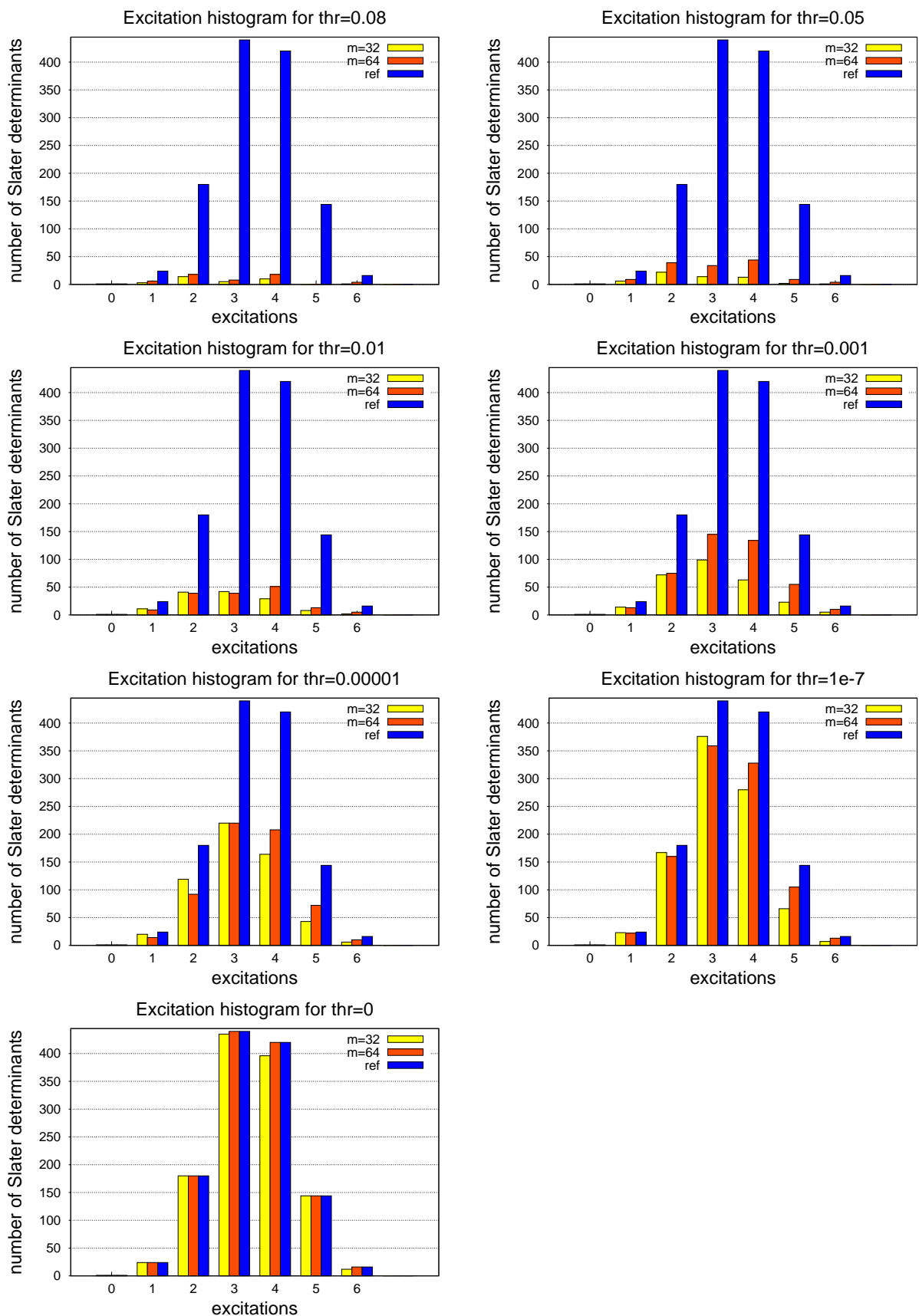}
\caption{Excitation histogram for CAS(7,7)CI-type wave function expansion reconstructed for DMRG(7,7)[$m$] calculations
with $m$ renormalized active-system states and for different threshold (thr) values of COM for the [Fe(NO)]$^{2+}$ molecule
surrounded by four point charges located at a distance of 1.131 \AA{} from the iron center. The CAS(7,7)SCF
reference excitation pattern is also shown and marked as ref.
}
\end{figure}

\clearpage
\begin{figure}[ht]
\centering
\includegraphics[width=0.6\linewidth]{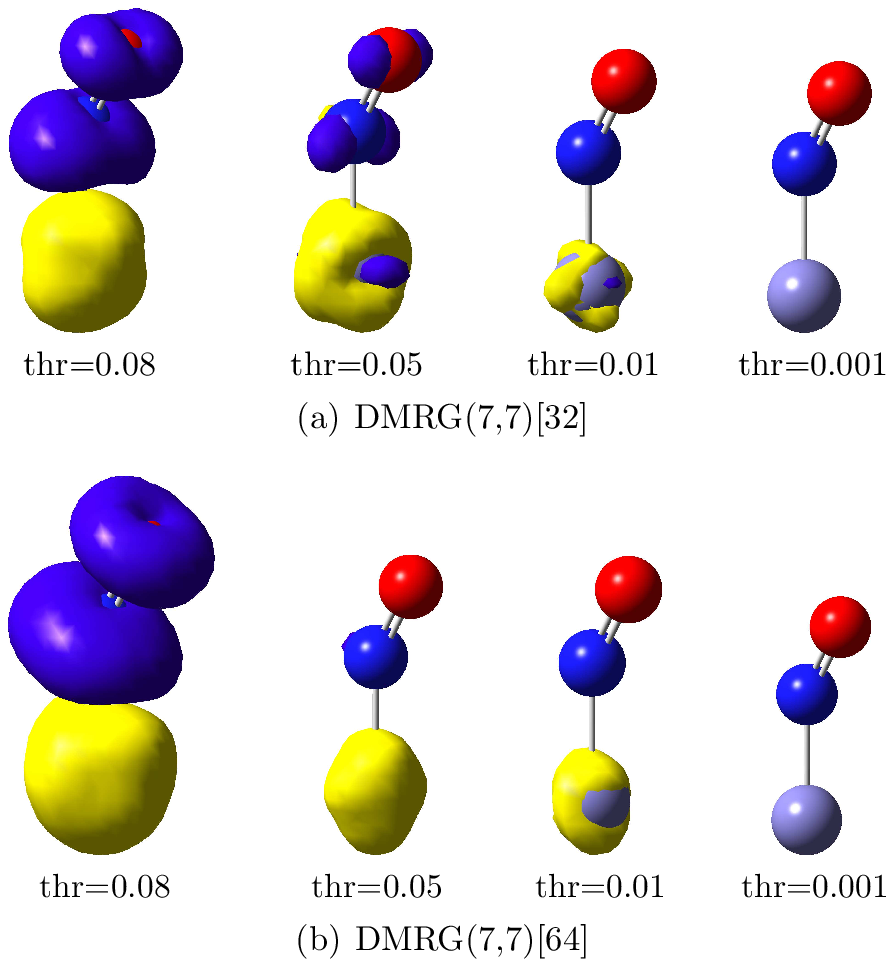}
\caption{CAS(7,7)CI[$m$]--DMRG(7,7)[$m$] spin density difference plots determined for different sampled subspaces of the
complete $N$-particle Hilbert space for the [Fe(NO)]$^{2+}$ molecule surrounded by four
point charges located at a distance of 0.598 \AA{} from the iron center. The spin density differences are plotted
for the reconstructed CAS(7,7)CI-type wave function with respect to the DMRG(7,7)[$m$] reference spin density for the
same number of renormalized active-system states $m$. The threshold (thr) indicates the upper bound of the COM value in
the sampling reconstruction procedure for each CASCI-type wave function. An isosurface value of 0.0003 was chosen
for all spin density difference plots.
}
\end{figure}

\begin{figure}[ht]
\centering
\includegraphics[width=0.8\linewidth]{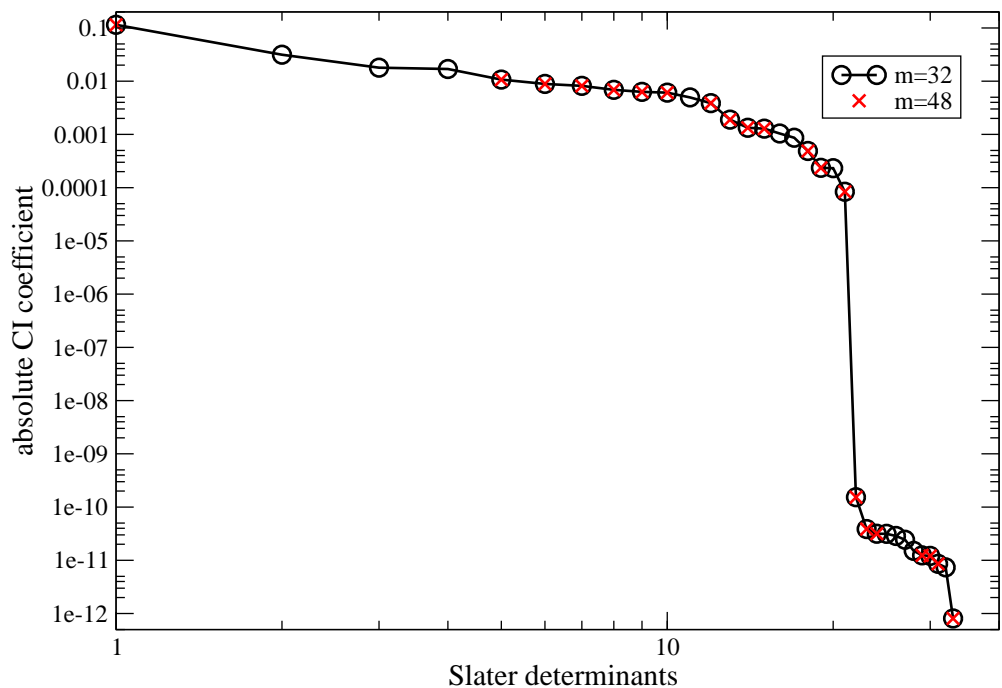}
\caption{Slater determinants which are not incorporated by the DMRG algorithm for different renormalized active-system
states $m$ in the DMRG(7,7)[$m$] calculations for [Fe(NO)]$^{2+}$ surrounded by four point charges in a distance of $d_{\rm pc} = 0.598$ \AA{} from the iron atom.
The missing Slater determinants are ordered according to their CI weight.
Increasing $m$, embeds some of the previously missing configurations. The
incorporated configurations for $m=48$ are displayed as open circles in the $m=32$ curve, while crossed circles
correspond to missing Slater determinants for both $m=32$ and $m=48$.
Note that important configurations which contain large CI weights could not be incorporated
by the DMRG algorithm in small-$m$ calculations.
}
\end{figure}
\clearpage
\newpage
%%%%%%%%%%%%%%%%%%%%%%%%%%%%%%%%%%%%%%%%%%%%%%%%%%%%%%%%%%%%%%%%%%%%%%%%%%%%%%%%%%%%%%%%%%%5
\section*{CASSCF and DMRG spin density convergence for intermediate CAS \label{subSec:results}}
%%%%%%%%%%%%%%%%%%%%%%%%%%%%%%%%%%%%%%%%%%%%%%%%%%%%%%%%%%%%%%%%%%%%%%%%%%%%%%%%%%%%%%%%%%%5
In Ref.~[36], different medium-sized active orbital spaces are discussed ranging from 11 electrons correlated in 9 orbitals up
to 13 electrons correlated in 16 active orbitals. In our DMRG study, we first focus on five different active orbital spaces, namely on
a CAS(11,9), CAS(11,11), CAS(11,12), CAS(11,13), and a CAS(11,14). In the CAS(11,9), the active orbital space is composed of
five Fe $3d$-orbitals ($d_{xy}$, $d_{xz}$, $d_{yz}$, $d_{z^2}$, and $d_{x^2-y^2}$), both NO $\pi$- and $\pi^*$-orbitals.
The latter four CASSCF calculations contain an additional shell of Fe $d$-orbitals: the Fe $d_{xz}$- and $d_{xy}$- (CAS(11,11)),
$d_{yz}$- (CAS(11,12)) and $d_{z^2}$-double-shell orbitals (CAS(11,14) which contains also an additional empty ligand orbital).
As orbital basis, natural orbitals from the corresponding CASSCF
calculations are employed [68-70]. All CASSCF calculations as well as the calculation of the one-electron and two-electron integrals
were performed with the \textsc{Molpro} program package [58] using Dunning's cc-pVTZ basis set
for all atoms [59,60].
As the convergence of DMRG calculations can crucially depend on the ordering of orbitals in the algorithm [Moritz, Hess, Reiher, {\it J. Chem. Phys.}
{\bf 2005}, {\it 122}, 024107], the DMRG orbital orderings were optimized by minimizing the entanglement among molecular orbitals
following the ideas of Refs.~[49,57,72]. 
Additionally, we applied the CI-DEAS procedure in the start up procedure in order to keep the system size small and accelerate convergence [57].
All DMRG calculations are performed with the Budapest DMRG program [71].
The number of DMRG active-system states $m$ ranges from 220 for the smallest active orbital space CAS(11,9) up to 1680 for the CAS(11,14) calculation
in order to reproduce the corresponding CASSCF reference energy, while for the intermediate active orbital spaces
$m=790$ (CAS(11,11)), $m=960$ (CAS(11,12)) and $m=1280$ (CAS(11,13)) active-system states, respectively, are required to reproduce the
CAS($x$,$y$)SCF energies and spin density distributions. 

For all these DMRG calculations, the spin density distribution of the corresponding
CASSCF reference calculation could be exactly reproduced and is therefore not displayed here. As an additional convergence
criterion for the DMRG calculations, the reconstructed CASCI-type wave function of each DMRG calculation can be analyzed and
compared to the exact CASSCF reference wave function. To perform such an analysis, we calculated the weighted quantum fidelity
for a reconstructed CASCI-type wave function and the corresponding CASSCF reference wave function considering only the most important configurations
(CI $> 10^{-6}$). The quantum fidelity was further normalized with respect to the norm of the truncated CASSCF reference wave function since all configurations with CI coefficients
smaller than $10^{-6}$ were neglected. The set of quantum fidelity measures $F_{\rm DMRG,CASSCF}$ for \{CAS(11,9),CAS(11,11),CAS(11,12),CAS(11,13),CAS(11,14)\}
is \{0.999999,0.999999,0.999999,0.999998,0.9999989\}.
Hence, the overlap of both wave functions is excellent, and similar spin
density distributions are obtained.

\begin{table}[Th]
\caption{The absolute error $\Delta_{\rm abs}$ and the root-square error $\Delta_{\rm rs}$ of DFT spin
densities with respect to the converged DMRG(13,29)[2048] reference spin density
for [FeNO]$^{2+}$ surrounded by four point charges at a distance of $d_{\rm pc}=1.131$ \AA{} from
the iron center.
}
\begin{center}
\begin{tabular}{lcc}\hline \hline
Method& $\Delta_{\rm abs}$ & $\Delta_{\rm rs}$ \\ \hline
 OLYP & 0.360309   &  0.0826634\\
 OPBE & 0.395280   &  0.0891197\\
 BP86 & 0.158527   &  0.0346259\\
 BLYP & 0.138796   &  0.0304860\\
 TPSS & 0.178942   &  0.0425946\\
 TPSSh& 0.718517   &  0.1634330\\
 M06-L& 0.446750   &  0.1048820\\
 B3LYP& 0.951955   &  0.2159410\\
\hline\hline
\end{tabular}
\end{center}
\end{table}

\end{document}